\documentclass[a4paper, 11pt]{article}

\usepackage[utf8]{inputenc}
\usepackage[OT1]{fontenc}
\usepackage[textwidth=15cm]{geometry}

\usepackage{graphicx}
\usepackage{amsmath}
\usepackage{amsfonts}
\usepackage{amssymb}
\usepackage{amsbsy}
\usepackage[export]{adjustbox}[2011/08/13]

\usepackage{latexsym}
\usepackage{array}

\usepackage{supertabular,booktabs,multirow,rotating}
\usepackage{hyperref}
\usepackage{color}
\definecolor{bluegray}{rgb}{0.0, 0.0, 0.7}
\hypersetup{
    colorlinks=true,
    linkcolor=bluegray,
    citecolor=bluegray,
    urlcolor=bluegray,
}
\usepackage{float}
\usepackage{textcomp}
\usepackage{times}
\usepackage{authblk}
\usepackage{caption} 
\usepackage{natbib}
\setcitestyle{authoryear,open={(},close={)}} 

\newcommand{\mb}{\mathbf}
\newcommand{\mbb}{\mathbb}

\newcommand{\bc}{\begin{center}}
\newcommand{\ec}{\end{center}}

\newcommand{\beq}{\begin{equation}}
\newcommand{\eeq}{\end{equation}}
\newcommand{\beqa}{\begin{eqnarray}}
\newcommand{\eeqa}{\end{eqnarray}}
\newcommand{\beqan}{\begin{eqnarray*}}
\newcommand{\eeqan}{\end{eqnarray*}}
\newcommand{\bit}{\begin{itemize}}
\newcommand{\eit}{\end{itemize}}

\newgeometry{top=2.5cm, bottom=2.8cm}

\begin{document}

\title{Data Integration with Fusion Searchlight: Classifying Brain States from Resting-state fMRI}

\author[1]{Simon Wein}
\author[1]{Marco Riebel}
\author[1]{Lisa-Marie Brunner}
\author[1]{Caroline Nothdurfter}
\author[1]{Rainer Rupprecht}
\author[1]{Jens V. Schwarzbach}

\affil[1]{\normalsize Department of Psychiatry and Psychotherapy, University of Regensburg, Regensburg, Germany}

\date{}

\maketitle


\begin{abstract}
Resting-state fMRI captures spontaneous neural activity characterized by complex spatiotemporal dynamics. Various metrics, such as local and global brain connectivity and low-frequency amplitude fluctuations, quantify distinct aspects of these dynamics. However, these measures are typically analyzed independently, overlooking their interrelations and potentially limiting analytical sensitivity. Here, we introduce the Fusion Searchlight (FuSL) framework, which integrates complementary information from multiple resting-state fMRI metrics. We demonstrate that combining these metrics enhances the accuracy of pharmacological treatment prediction from rs-fMRI data, enabling the identification of additional brain regions affected by sedation with alprazolam. Furthermore, we leverage explainable AI to delineate the differential contributions of each metric, which additionally improves spatial specificity of the searchlight analysis. Moreover, this framework can be adapted to combine information across imaging modalities or experimental conditions, providing a versatile and interpretable tool for data fusion in neuroimaging.
\end{abstract}

\section{Introduction}

Complex cognition arises from the interplay of activity among billions of neurons. A fundamental goal in studying neurological diseases and mental health disorders is to understand how these interactions may be systematically altered to explain behavioral changes. One promising noninvasive approach involves studying the brain at rest \citep{Biswal1995_rs} and characterizing brain activity using measures of local BOLD activity, as well as localized and distributed connectivity metrics. These metrics can be compared between clinical groups and healthy controls or between a baseline and a post-intervention state. 

\paragraph{Resting-state fMRI markers}
Functional connectivity (FC) is one popular approach for studying global interactions between brain regions from a functional network perspective \citep{Vandenheuvel2010_rs_review_fc}. Thereby, measures from graph theory allow us to detect changes in connectivity properties of brain networks \citep{Bassett2017_networks}. On a smaller scale, regional homogeneity (ReHo) quantifies local connectivity across the cortex in the range of millimeters, and is defined as the temporal coherence or synchrony of the BOLD signal of neighboring voxels or vertices \citep{Jiang2016}. In addition to connectivity measures, the fractional amplitude of low frequency fluctuations (fALFF) can characterize local modulations in the amplitude of signals \citep{Yang2007, Zou2008}. In this manner, these different resting-state fMRI (rs-fMRI) metrics capture complementary aspects of BOLD signal dynamics \citep{Lv2018_review_rs-fMRI}. 

\paragraph{In clinical rs-fMRI studies different markers can be modulated simultaneously}
In rs-fMRI studies on depression, alterations have been linked to modulations in FC \citep{Javaheripour2021_FC_depression, Kaiser2015_FC_depression}, fALFF \citep{Gao2021_falff_depression, Wang2022_falff_depression} and ReHo \citep{Shen2017_reho_depression, Ni2023_reho_depression, Sun2022_reho_depression}. Also multiple sclerosis has shown to affect a multitude of markers, such as FC \citep{Tahedl2018_review_fc_ms, Huang2020_ms_fc, Tahedl2022_fc_ms}, fALFF \citep{Liu2015_ms_falff} and ReHo \citep{Wu2016_ms_reho}. 
Furthermore, sedative related effects have shown to be apparent across multiple different rs-fMRI metrics. Administering chloral hydrate, dexmedetomidine and propofol resulted in decreases in whole-brain FC density and efficiency \citep{Wei2013_connectivity_ch, Hashmi2017_dexm_sedation, Schrouff2011_propofol_wb_conn}. Intake of alprazolam increased ReHo in sensory- and sensory-integration areas \citep{Hinojo2021_alprazolam}. Administering diazepam \citep{pflanz_effects_2015} and midazolam \citep{Liang2015_midazolam_ICA} yielded
an increase of within-network connectivity mainly in low-level sensory networks. Furthermore, midazolam elevated low frequency amplitudes ($ < 0.05 Hz $) in the visual, sensorimotor and auditory resting-state network \citep{Kiviniemi2003, Kiviniemi2005}.
However, many current rs-fMRI studies focus only on a single metric, or analyze these biomarkers separately. In this manner, information on important modulations of the BOLD signal may be neglected, potentially reducing sensitivity of detecting altered brain states\footnote{Based on the definition of \cite{Greene2023_brain_state}, we refer as brain states to recurring activity patterns distributed across the brain that emerge from physiological or cognitive processes.}.

\paragraph{Integrating rs-fMRI metrics with fusion searchlight} 
In our study we present an interpretable framework based on searchlight (SL) decoding for fusing such complementary fMRI markers and improving analysis sensitivity. SL decoding uses multivariate pattern analysis (MVPA) to test the information content in small spherical subsets in every voxel or vertex across the cortex \citep{Kriegeskorte2006_sl, Etzel2013_pitfalls}. In this manner SL typically produces spatial maps of decoding accuracies, highlighting areas that are informative for a specific brain state. In our fusion SL (FuSL) framework we integrate the distinct information of BOLD dynamics expressed in local ReHo, fALFF and global FC and show how this increases decoding performance in rs-fMRI. In a next step we utilize an explainable AI (XAI) approach to retrospectively trace back the feature importance of each metric for decoding. To this aim, we rely on Shapley additive explanations (SHAP) \citep{Lundberg2017_shap}, a XAI method based on game theoretically optimal Shapley values \citep{Shapley1953}. This explanation method has shown to combine several desirable mathematical properties such as \textit{local accuracy}, \textit{missingness} and \textit{consistency}. Furthermore this model-agnostic explanation approach allows us to use any MVPA decoding model in our workflow.

\paragraph{Increase in spatial specificity}
We show that incorporating XAI to evaluate feature importance has the additional benefit of increasing spatial specificity in SL analysis. Solely analyzing prediction accuracy, as typically done in current studies applying SL decoding, only reveals if there is somewhere within the SL radius information present. Thus, maps of decoding accuracies do not exactly represent maps of informative voxels \citep{Etzel2013_pitfalls}. Incorporating SHAP as a post hoc test can help us to reconstruct the precise locations of voxels within a SL that are most important for decoding, what helps us to increase spatial specificity of SL analysis.

\paragraph{Experiments}
We first investigate general properties of the FuSL framework employing an artificial MRI dataset and study how different combinations of informative and uninformative data sources affect its decoding performance at different levels of noise, levels of spatial autocorrelations and levels signal inhomogeneities across samples. We further demonstrate that the integration of FC, ReHo and fALFF can enhance the prediction of pharmacological treatment of participants with rs-fMRI data from a pharmacological fMRI study \citep{Wein2024_npp}.
Thereby, the increase in decoding accuracy by integrating information across metrics allows us to identify additional regions that are significantly altered by treatment with the benzodiazepine alprazolam. Finally we show how SHAP can then be used to trace back the differential impact of alprazolam on the individual fMRI markers. Based on these applications we demonstrate how FuSL can provide a flexible and interpretable framework for data fusion in neuroimaging.

\section{Materials and Methods}

\subsection{Datasets}

\subsubsection{Artificial dataset}

In the beginning we use an artificial dataset to investigate basic properties of FuSL when combining different data sources. On one side we want to study decoding performance of FuSL when signal or information is present in multiple imaging metrics or modalities simultaneously in the same location. On the other side we explore the behavior when we combine noisy non-informative data sources with informative ones in our FuSL. To study these scenarios we generate 3 artificial source signals, which could represent 3 different neuroimaging metrics or modalities. In this toy example the objective of the FuSL is, based on these 3 sources, to discriminate between two groups. We generated data for two groups, each containing 30 samples, by inducing artificial signals in one group, but not the other. Therefore we set signal values to 1 within a circular region of interest (ROI). In source 1 and 2 we simultaneously induce a signal in the left frontal cortex setting values within a 5 mm radius to 1 and -1 respectively (defined as ROI 1). As a third source we induced a signal in the left parietal cortex, setting values to 1 within a 5 mm radius (defined as ROI 2). We then generated 30 samples from these signals by sampling the signal amplitudes from a Gaussian normal distribution with standard deviations of $\sigma = 1$ and  $\sigma = 2$, to obtain datasets with different amount of signal variability across samples, as typically observed in different clinical or experimental studies. We added random Gaussian noise to each sample with varying powers $ P_{noise} = [1, 10, 20]$ resulting in datasets with different signal to noise ratios $S/N = [1, 0.1, 0.05]$ respectively. In this manner we created datasets for which signals in source 1 and 2 overlapped in ROI 1, and one signal was only expressed in ROI 2 in source 3. This allowed us to study in ROI 1 the impact of combining informative data sources, and in ROI 2 the impact of adding uninformative sources to informative ones. To further address the diversity in preprocessing protocols and modalities of data possibly integrated in FuSL, we additionally studied the impact of increasing the amount of spatial autocorrelations in artificial data. For this aim, we smoothed the data with a Gaussian filter with different full widths at half maximum $ FWHM = [2 mm, 4 mm, 6 mm] $ and evaluated decoding performance in dependency of the spatial smoothness. Finally, we also conducted a power analysis by sub-sampling from the whole artificial dataset and evaluated the proportion of vertices that exceeded statistical significance in dependence of the sample size.

\subsubsection{Resting-state fMRI dataset}

\paragraph{Study description}

We used rs-fMRI data from 34 healthy male participants between the ages of 20 and 50 (mean = 27.2, std = 6.84) years and a BMI between 20.0 and 32.1  (mean = 24.3, std = 3.1). After screening for the absence of physical and psychiatric disorders all participants underwent rs-fMRI after a five-days intake of either alprazolam or placebo \citep{Wein2024_npp}. Participants gave their informed written consent at the beginning of the experiment and the study was approved by the local ethics committee (approval number 18-1197-111) and the National Institute for Pharmaceutical Security (BfArM, Eudra-CT-number: 2018-002181-40). Subjects received alprazolam (1.5 mg/d in 3 doses of 0.5 mg) or placebo for 5 days, each with at least 7 days washout phases between treatments.

\paragraph{MRI acquisition}

Resting-state fMRI data were collected with a Siemens Magnetom Prisma 3T Scanner at the University of Regensburg. Participants were instructed to relax and stay awake during the rs-fMRI session, while keeping their eyes closed. Based on pre-studies on optimizing the signal to noise ratio of echoplanar imaging (EPI) protocols \citep{seidel2020temporal}, we used an EPI multi-band sequence (multi-band factor 4) with a repetition time (TR) of 1000 ms, echo time (TE) of 30 ms and a flip angle (FA) of 60$^{\circ}$. During a scanning time of 22 min in total 1320 volumes were collected, with a field of view (FoV) of 192 mm $ \times $ 192 mm, an acquisition matrix (AM) of 64 $ \times $ 64, and an isotropic voxel size of 3 mm.
Additionally field map images were collected by using a double-echo spoiled gradient echo sequence, with TR = 715 ms, TE = 5.81/8.27 ms, FA = 40$^{\circ}$, with an isotropic voxel size of 3 mm, generating a magnitude image and two phase images. 
High resolution T1-weighted images were acquired using a Magnetization Prepared Rapid Gradient Echo (MP-RAGE) sequence, with a TR = 1919 ms, TE = 3.67 ms, FA = 9$^{\circ}$, AM = 256 $ \times $ 256 and FoV = 250 mm $ \times $ 250 mm.

\paragraph{Preprocessing}

For preprocessing the functional and structural imaging data, we incorporated the processing pipeline provided by fMRIPrep{\footnote{\url{https://fmriprep.org/en/stable/}}} (version 20.2.4) \citep{Oscar2017_fmriprep}. Structural preprocessing included averaging T1w images across sessions, bias field correction based on Advanced Normalization Tool's (ANTs) `N4BiasFieldCorrection' \citep{Tustison2010_ants}, brain tissue segmentation using FSL's `fast' \citep{Jenkinson2012_fsl}, brain surface reconstruction using FreeSurfer's `recon-all' \citep{Fischl2012_fs} and spatial normalization to the MNI152 standard space using ANT's `antsRegistration' \citep{Avants2011_ants_reg}. Functional preprocessing included susceptibility distortion correction based the double-echo gradient field maps using fMRIPrep's costum workflows \citep{Oscar2017_fmriprep}. Functional images were registered to the anatomical volume using Freesurfer's `bbregister'. Head motions were estimated usings FSL's `flirt' and slice time correction was applied using AFNI's `3dTshift' \citep{Cox1996_afni}. Finally volumetric fMRI data were projected onto the cortical surface, generating high-resolution grayordinate time courses in the fsLR standard space \citep{Glasser2013}. Prior to the ReHo and fALFF analysis we smoothed the fMRI data using Gaussian surface smoothing with a FWHM of 3 mm \citep{Glasser2013}.

\paragraph{Resting-state fMRI metrics}

Different rs-fMRI metrics allowed us to capture the different characteristics of the complex spatio-temporal dynamics of the BOLD signal.
For analyzing local connectivity based on regional homogeneity (ReHo), we selected for each vertex its 5-hop neighboring vertices in the fsLR standard space \citep{Glasser2013}, and computed ReHo as the average Pearson correlation between neighbouring BOLD time courses within the 0.01 Hz - 0.1 Hz frequency range. 
Besides local connectivity, fractional amplitude of low frequency fluctuations (fALFF) can supplement such an analysis by detecting local modulations in the signals amplitude \citep{Yang2007, Zou2008}. This quantity reflects the ratio of the BOLD signal's power spectrum of low frequencies to the signal's entire frequency range. In our study we focused on the very-low frequency range 0.01 Hz - 0.05 Hz, which has been reported to increase during midazolam sedation and anesthesia \citep{Kiviniemi2000, Kiviniemi2003, Kiviniemi2005}.
In addition to these local metrics, whole-brain functional connectivity (FC) allowed us to investigate modulations in brain networks from a more global perspective. Measures from graph theory thereby helped us to characterize changes in connectivity properties of individual regions of interest (ROIs).
To evaluate whole-brain FC we subdivided each hemisphere into $ 180 $ ROIs defined by \cite{Glasser2016}. We computed the average activity time course within each ROI and filtered the resting-state BOLD signal within the 0.01 Hz - 0.1 Hz low frequency range. FC was then defined as the Pearson correlation coefficient $ r $ between time courses of pairs of ROIs to obtain a $ 360 \times 360 $ whole-brain FC matrix. We selected a moderate FC threshold $ r \geq 0.6 $ yielding an average connection density of $ 39.2 \% $ and analyzed modulations in functional connectivity efficiency (FCE) for each ROI in the network.
According to the weights of the probabilistic multi-modal parcellation \citep{Glasser2016}, for each vertex we computed FCE as a weighted sum of the ROI-wise FCE values, weighting each vertex with the probability to belong to a respective ROI.
This yielded a spatially continuous distribution of FCE values across vertices. A mathematical definition of these three metrics is additionally provided in supplement I. We computed each of these metrics for the alprazolam and placebo condition of each subject.

\subsection{Fusion searchlight}

\begin{figure}[htbp]\begin{center}\includegraphics[width=0.99\textwidth]{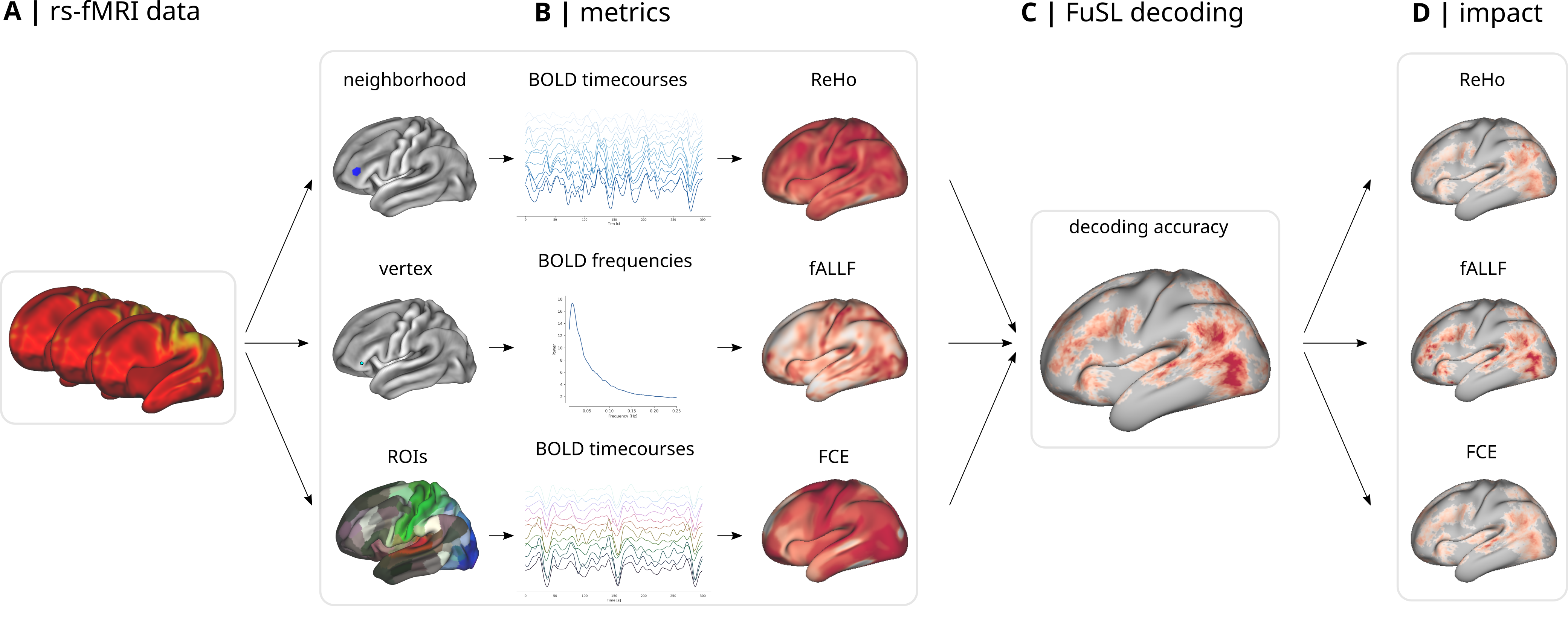}
\caption{Individual steps of the FuSL framework. (A) In resting-state fMRI we can observe a BOLD signal exhibiting complex spatial and temporal dynamics. (B) Using different metrics we can characterize different aspects of these neural dynamics. Regional homogeneity (ReHo) quantifies the coherence of the BOLD signal within local neighborhoods across the cortex. Fractional amplitudes of low frequency fluctuations (fALLF) are defined as the ratio of the low-frequency power to the power of the entire frequency range for each vertex. Functional connectivity can be used to characterize functional connection efficiency (FCE) of individual ROIs in the global brain network. (C) In our FuSL framework these different metrics are concatenated to identify brain regions that are informative for decoding a brain state. (D) We used Shapley additive explanations (SHAP) to retrospectively reconstruct the impact of each metric on the decoding at a specific location.}
\label{fig:FuSL_workflow}\end{center}\end{figure}

In our FuSL framework, we integrated the complementary information provided by different data sources in a searchlight-based decoding analysis. For decoding pharmacological treatment of subjects we computed ReHo, fALFF and FCE at each vertex, and concatenated these values within each SL to train a classifier (figure \ref{fig:FuSL_workflow} A and B). In this repeated measurement fMRI study we removed the within subject mean of each rs-fMRI metric to account for the treatment independent variability of metrics across subjects. In consideration of the variability in density of the fsLR standard space mesh  (illustrated in supplementary figure S1), we defined SLs based on a k-hop neighborhood radius to ensure a constant number of features used to train each SL decoder across the cortex. 
We used cross validation (CV) with stratification to balance the classes within each split \citep{Kohavi1995}. To keep the test set completely independent we held out both fMRI sessions of each subject for testing at each CV split. After training and testing the SL classifier we computed a spatial map of test accuracies averaged across folds, pointing out vertices with neighborhoods that are informative to decode a specific brain state, like in classical SL analysis (figure \ref{fig:FuSL_workflow} (C)). To identify test accuracies that were significantly different from chance level we permuted the labels 1000 times and applied threshold-free cluster enhancement (TFCE) to the original and permuted test accuracy maps \citep{Smith2009_tfce}. To correct for multiple comparisons across vertices we took the maximum TFCE values of test accuracy maps derived from permuted data to build the null distribution and tested the values of the original TFCE map against them \citep{Smith2009_tfce}. For the statistical testing of the decoding accuracy of different models across CV folds, we employed a paired t-test incorporating the variance correction proposed by \cite{Nadeu1999_ttestcv} to account for correlations in the training data across folds \citep{Bouckaert2004_reliability_CV}. We applied false discovery rate (FDR) correction to account for multiple comparisons of different models \citep{Benjamini1995_FDR}. For the statistical analysis we utilized functions provided by the SciPy{\footnote{\url{https://scipy.org/}}} and statsmodels{\footnote{\url{https://www.statsmodels.org/stable/index.html}}} Python packages \citep{Virtanen2020_SciPy, Seabold2010_statsmodels}.

In a next step we aimed to explain the predictions of the FuSL classifier. To reveal which metrics drove the decision of the SL, we used Shapley additive explanations (SHAP) to reconstruct the importance of each metric at each location on the cortex \citep{Lundberg2017_shap}. This explanation approach assigns post-hoc an importance score to each input feature of a classifier, which we utilized to determine whether modulations in ReHo, fALFF or FCE were indicative for decoding a specific brain state. For each SL we first computed the feature importance of each metric at each vertex and then averaged these values across overlapping SLs at each vertex. In this manner we obtained one importance score for each metric at each vertex on the cortex. A formal derivation of these importance measure is provided in the supplementary material. To obtain maps that reflect simultaneously the \textit{importance} and the \textit{informativeness} of a metric, we further weighted spatial maps of SHAP values (feature importance) with the spatial map of decoding test accuracies (informativeness). We denoted this combination of informativeness and importance as the actual \textit{impact} of a metric (figure \ref{fig:FuSL_workflow} (D)).
However, these maps do not provide us with information whether a metric increased or decreased due to an intervention or condition, which can be valuable information for the interpretation of analysis results. To address this shortcoming we propose to additionally incorporate a third characteristic of the data. We computed the differences in metrics between the groups across the cortex to obtain a spatial map indicating the intervention-associated changes of each metric. Finally we weighted these maps of feature differences with the feature impact. This yielded spatial maps that condensed the modulation, importance and informativeness of each metric at each location for a compact visual interpretation. We denote this combination as the \textit{feature-weighted} impact. For a more specific analysis of these three aspects one can also visualize these maps individually, depending on the requirements of MRI study. An implementation including a demo version is available at: \url{https://github.com/simonvino/FuSL} which is based on functionalities provided by NiLearn{\footnote{\url{https://nilearn.github.io/stable/index.html}}} \citep{Abraham2014_nilearn}, scikit-learn{\footnote{\url{https://scikit-learn.org/stable/}}} \citep{Pedregosa2011_sklearn}, and the SHAP software package{\footnote{\url{https://shap.readthedocs.io/en/latest/}}} \citep{Lundberg2017_shap}.

\section{Results}

\subsection{Artificial dataset}

\begin{figure}[htbp]\begin{center}\includegraphics[width=0.85\textwidth]{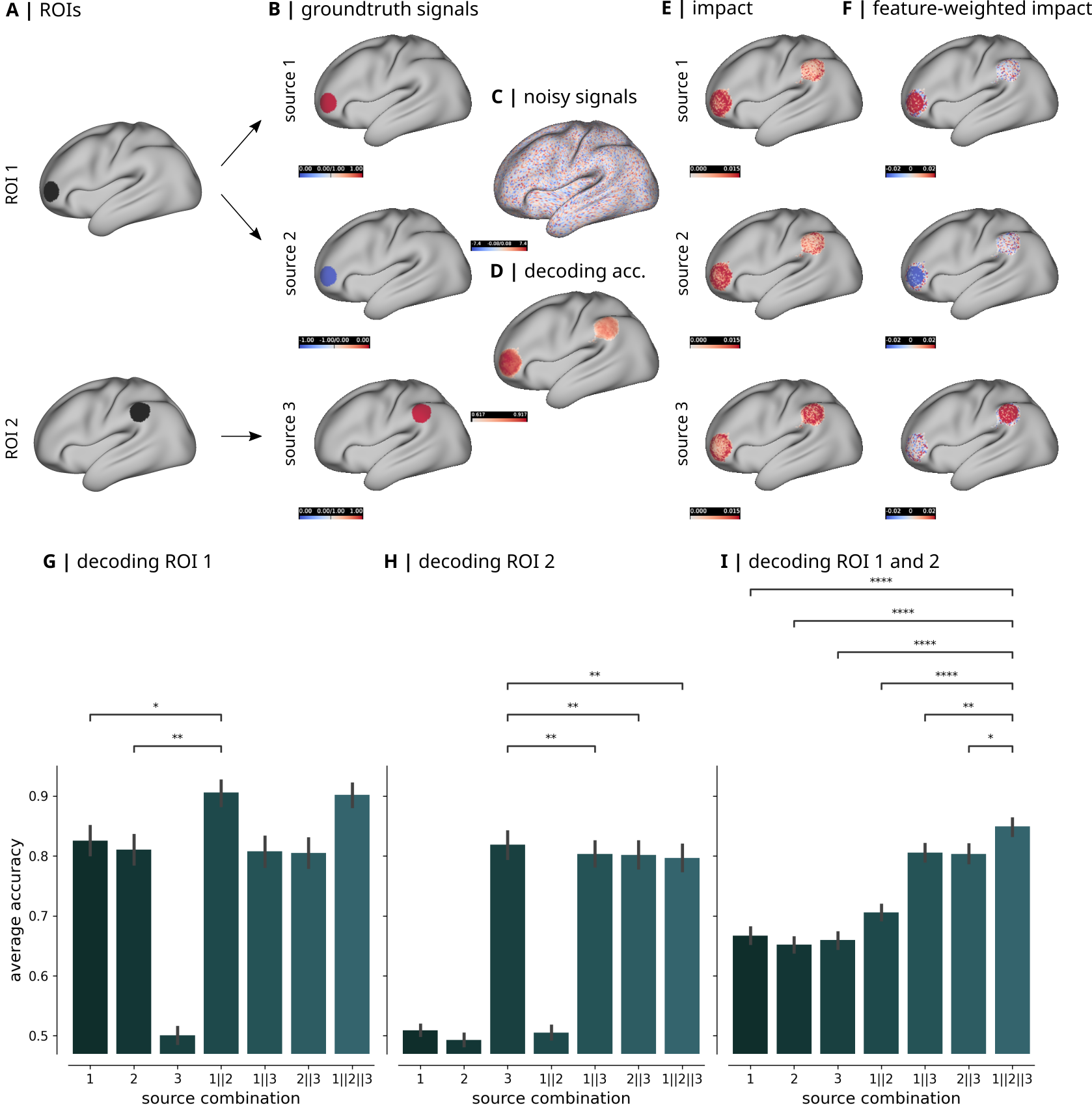}
\caption{Investigating basic properties of the FuSL framework using an artificial dataset. We induce artificial signals in two ROIs (A), whereby signals of source 1 and 2 overlap in ROI 1 and signals in source 3 are only present in ROI 2 (B). (C) We then generate samples by randomly varying the signal amplitude and adding Gaussian noise. (D) We observed higher decoding test accuracy values in ROI 1 than in ROI 2. (E) Impact values of source 1 and 2 are higher in ROI 1 than ROI 2, and of source 3 higher in ROI 2 than ROI 1. (F) The feature-weighted impact displays increased values in source 1 and 3 and the decrease of values in source 2. (G) and (H) show decoding accuracies for all combinations of sources in ROI 1 and 2 respectively. (I) Decoding accuracies averaged across ROI 1 and 2. Error bars represent 95\% confidence intervals across folds. Significant differences of accuracies in figure (G), (H) and (I) are indicated with: *: $ p \leq 0.05 $, **: $ p \leq 0.01 $, ***: $ p \leq 0.001 $, ****: $ p \leq 0.0001 $, ns: not significant, (ns): not significant after false discovery rate correction.}
\label{fig:arti_data}\end{center}\end{figure}

We first investigated basic properties of the FuSL framework using the artificial dataset with a moderate signal to noise ratio of $S/N = 1$, a signal variability of $\sigma = 1$ across samples and with no spatial smoothing. Figure \ref{fig:arti_data} (A) illustrates the 2 ROIs with artificial signals, whereby signal in ROI 1 was observed in sources 1 and 2, whereas the signal in ROI 2 was only observed in source 3, as depicted in figure \ref{fig:arti_data} (B). 
An example of one sample with a noisy signal is shown in figure \ref{fig:arti_data} (C). We then studied the decoding accuracy using different combinations of these sources as inputs for the FuSL. To fit the FuSL we used 10-fold stratified CV with 10 repetitions and used a SL radius of 3 vertices. As a decoder we employed a support vector classifier (SVC) with a radial basis function (RBF) kernel \citep{Cortes1995_SVM, Schoelkopf1998_kernels}. Figure \ref{fig:arti_data} (D) shows statistically significant decoding test accuracies across the cortex when one incorporates all three sources simultaneously. In ROI 1 we observed higher decoding accuracy values than in ROI 2, because in ROI 1 the FuSL classifier could combine information simultaneously present in source 1 and 2. We then reconstructed the impact, as well as the feature-weighted impact of each source, as shown in figure \ref{fig:arti_data} (E) and (F). Impact values of source 1 and 2 were higher in ROI 1 than in ROI 2, and impact of source 3 was higher in ROI 2 than in ROI 1. Further we can see that feature-weighted impact maps reflected well the ground truth signals shown in figure \ref{fig:arti_data} (B), whereby we observed some variability due to the added noise. Incorporating differences in features between groups in these maps also allowed us to reconstruct the change of the signal in each source. In this manner, we were able to retrieve the increase of the signal in source 1, and a decrease of the signal in source 2.

\begin{figure}[htbp]\begin{center}\includegraphics[width=0.4\textwidth]{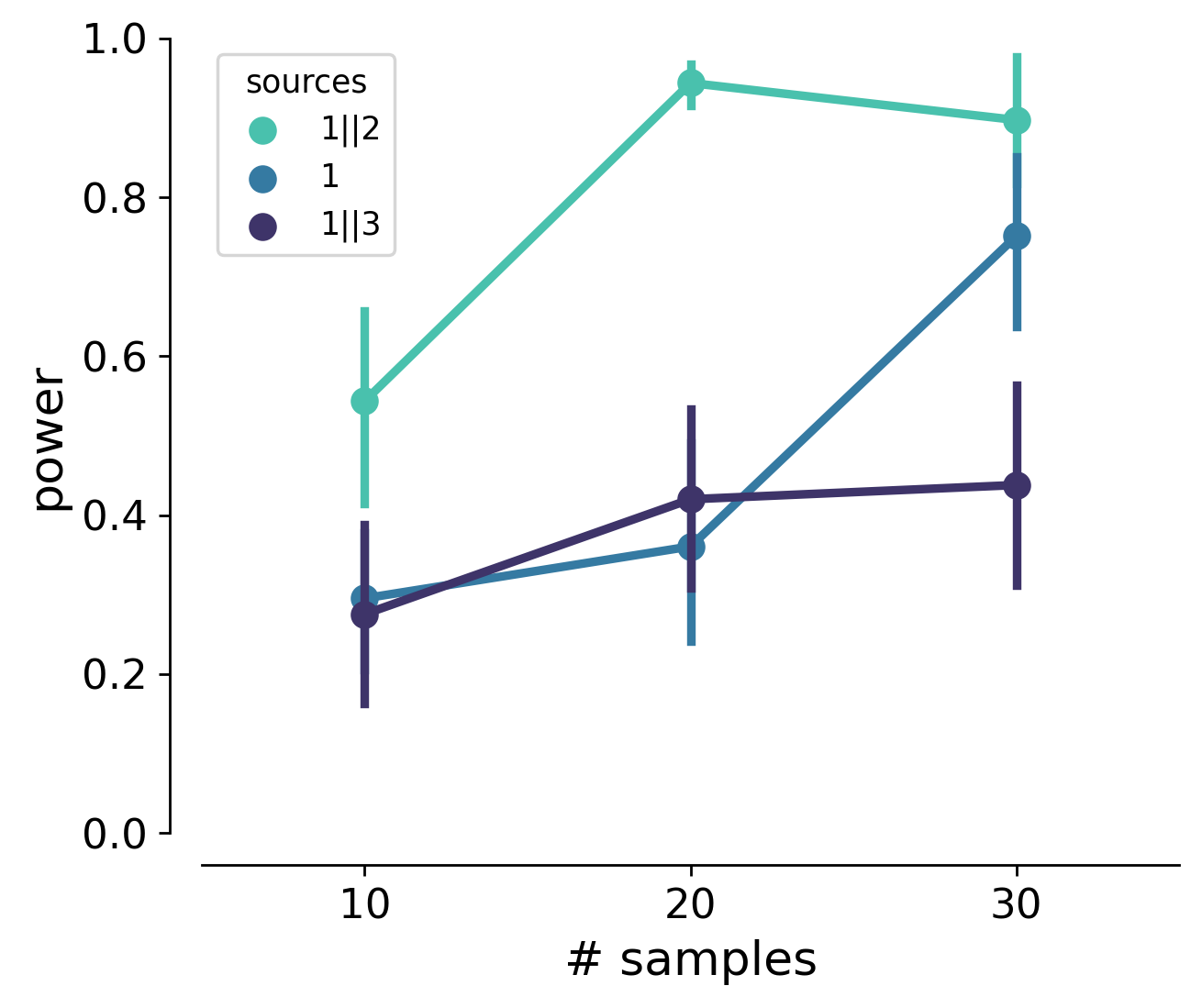}
\caption{Statistical power of the decoding analysis in ROI 1. Adding the informative source 2 to source 1 increased the power, especially when the sample size is smaller, whereas adding the uninformative source 3 led to a decrease in statistical power. Errorbars represent 95\% confidence intervals across dataset subsampling repetitions.}
\label{fig:arti_power}\end{center}\end{figure}

We then studied the decoding accuracy under different combinations of informative and uninformative data sources as inputs for FuSL. 
Figure \ref{fig:arti_data} (G) shows the average test accuracies in ROI 1 for all combinations of the three modalities, using 10 repetitions of 10-fold CV. We defined a combination of two sources using the $||$ operator, indicating a concatenation of their features. Using only source 1 or source 2 as input yielded an accuracy of $ 82.5 \% $ and $ 81.1 \% $ respectively, whereby using their combination 1$||$2 in FuSL increased the accuracy to $ 90.6 \% $ (Cohen's $d = 0.856$ and $d = 0.729$ respectively). 
In ROI 2 an average accuracy of $ 81.9 \% $ was achieved when using only the informative source 3 as an input, as shown in figure \ref{fig:arti_data} (H). The performance decreased slightly but significantly to $ 80.3 \% $ and $ 80.2 \% $ when noisy, uninformative sources were added (Cohen's $d = 0.138$ and $d = 0.150$ respectively), and to $ 79.7 \% $ when two uninformative sources were added (Cohen's $d = 0.200$). Figure \ref{fig:arti_data} (I) shows the average decoding accuracy of ROI 1 and 2. We observed that a combination of all sources 1$||$2$||$3 achieved the overall best decoding accuracy. The increase in decoding accuracy by adding an informative source was thereby considerably larger than the decrease when adding an uninformative source.
Figure \ref{fig:arti_power} shows the statistical power of the decoding in ROI 1 in dependency of the sample size. Adding the informative source 2 to source 1  considerably increased the statistical power, especially when sample sizes were smaller. Adding an uninformative source 3 to source 1 again led to a slight decrease in statistical power.

In addition we compared the performance of the SVC to different MVPA decoding models such as a k-nearest neighbors classifier, a random forest classifier and ridge regression classifier in figure S2 (A). We also studied the impact of different kernel functions for SVC, and we found that a SVC with RBF kernel overall yielded the best results. A comparison of SL radii in figure S2 (B) revealed that a neighborhood order of 3 provided a good tradeoff between decoding accuracy and computational time. 
A more comprehensive evaluation of the impact of data fusion across datasets with different signal to noise ratios ($S/N = [1, 0.1, 0.05]$), different amounts of spatial autocorrelation (smoothing with $ FWHM = [2 mm, 4 mm, 6 mm] $), and different signal standard deviations across samples ($\sigma = [1, 2]$) is additionally provided in figure S3 and S4. Spatial smoothing showed to especially improve decoding accuracy when noise levels are higher, but the overall characteristic accuracy gain when combining informative data sources was preserved across all datasets, demonstrating a good reliability of FuSL across different scenarios.

\begin{figure}[htbp]\begin{center}\includegraphics[width=0.7\textwidth]{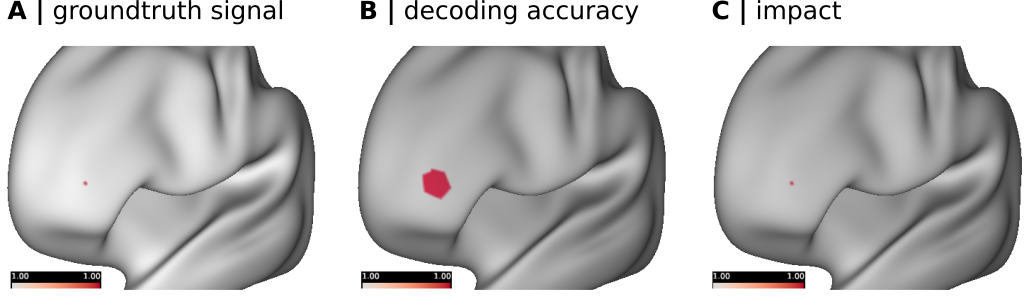}
\caption{Increasing spatial specificity of searchlight decoding using Shapley values. (A) We induce a signal in one vertex only. (B) The area of significant decoding accuracy is considerably larger than the original signal. (C) Analyzing the feature impact maps allows us to reconstruct again the exact location of an informative vertex.}
\label{fig:arti_specificity}\end{center}\end{figure}

We further show in an example, how reconstructing feature importance using Shapley values increased also the spatial specificity of detected regions that were relevant for decoding a brain state. To this aim, we generated a signal in one vertex on the surface as shown in figure \ref{fig:arti_specificity} (A). Current applications typically report the entire area of SLs with significant decoding accuracy \citep{Etzel2013_pitfalls, Viswanathan2012_sl_geometry}, which is shown in our example in figure \ref{fig:arti_specificity} (B). Such an area can be considerably larger than the area of the actual signal, because all SLs that contain the informative vertex may be able to successfully decode. Computing feature importance on vertex-level based on Shapley values allowed us to trace back the exact location of an important vertex within a SL. After averaging importance values across overlapping SLs, and weighting these maps with the decoding accuracy, the map of feature impact precisely identified the original location of the signal, as shown in figure \ref{fig:arti_specificity} (C).

\subsection{Resting-state fMRI dataset}

\begin{figure}[htbp]\begin{center}\includegraphics[width=0.87\textwidth]{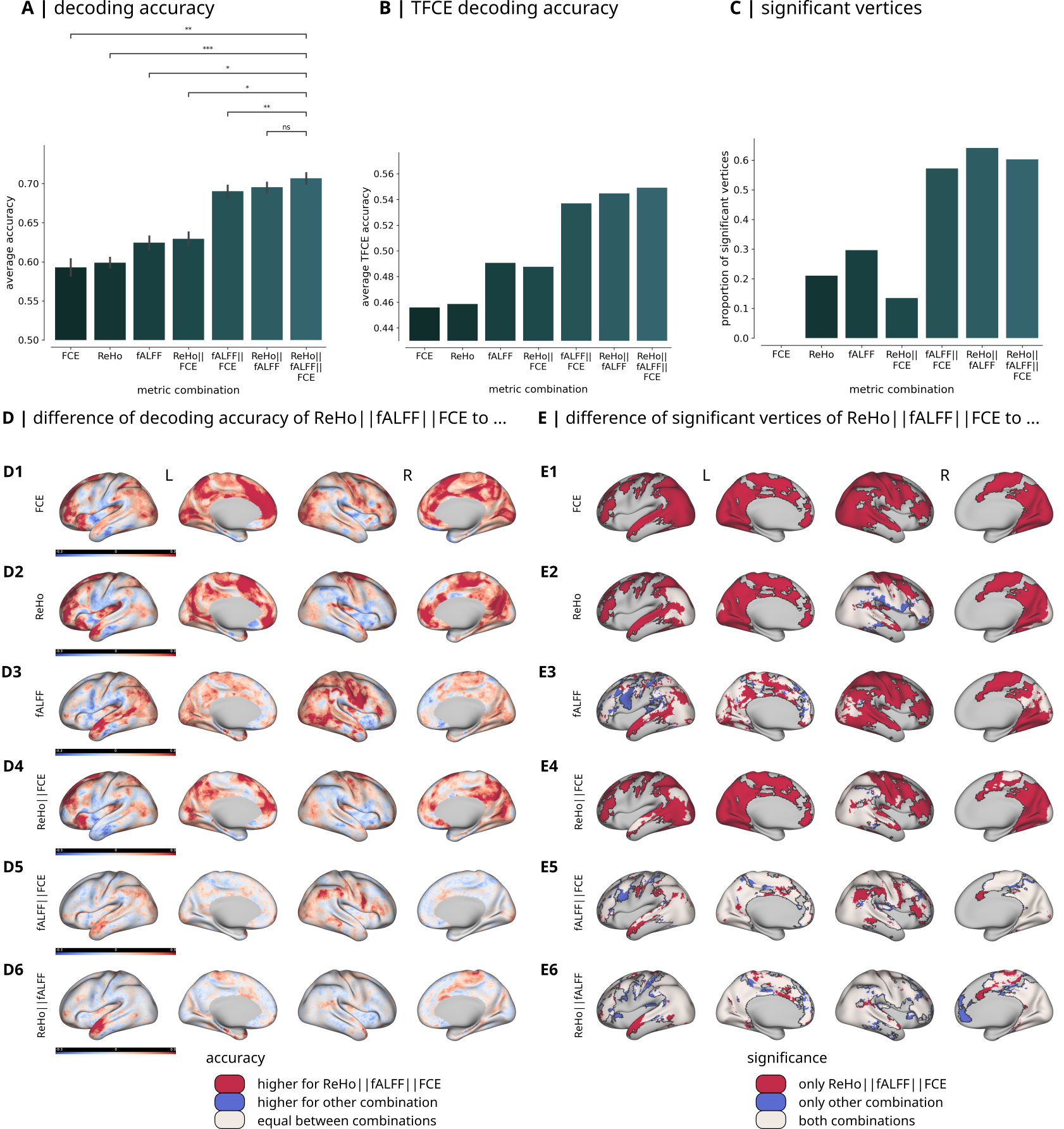}
\caption{Decoding performance of FuSL in dependency of rs-fMRI metric combinations. 
The combinations fALFF$||$FCE, fALFF$||$ReHo and ReHo$||$fALFF$||$FCE achieved highest average decoding test accuracies (A), highest threshold-free cluster enhanced accuracy (B) and largest number of significant vertices (C). 
(D) Local differences in decoding test accuracies between ReHo$||$fALFF$||$FCE and all other combinations in the left (L) and right (R) cortex. (E) Locations where only ReHo$||$fALFF$||$FCE yielded significant decoding accuracies are highlighted in red. Locations where only the respective other combination is able to decode are depicted in blue. Locations where ReHo$||$fALFF$||$FCE and the respective other combination can both significantly decode are marked in white. Significant differences of accuracies in figure (A) are indicated with: *: $ p \leq 0.05 $, **: $ p \leq 0.01 $, ***: $ p \leq 0.001 $, ns: not significant.}
\label{fig:tspo_comparison}\end{center}\end{figure}

In this application we studied how integrating complementary rs-fMRI metrics in FuSL could improve the prediction of pharmacological treatment in rs-fMRI data. We incorporated combinations of regional homogeneity (ReHo), fractional amplitude of low frequency fluctuations (fALFF) and functional connectivity efficiency (FCE) with the objective to predict whether participants have received the sedative alprazolam or a placebo. 
Figure \ref{fig:tspo_comparison} shows the performance of FuSL related to different combinations of rs-fMRI metrics. The decoding test accuracies averaged across the entire cortex and across 100 repetitions of 10-fold CV are depicted in figure \ref{fig:tspo_comparison} (A). Statistically significant differences of the average test accuracies are highlighted in figure \ref{fig:tspo_comparison} (A) and the respective p-values and effect sizes are provided in supplementary table S1. We further evaluated the number of vertices in which FuSL was able to find significant differences between groups, using 10-fold CV. 
The average TFCE weighted  test accuracy values and the proportion of significant vertices are shown in figure \ref{fig:tspo_comparison} (B) and (C) respectively. 
We found that mainly combining the information in fALFF with a connectivity based rs-fMRI metrics (ReHo, FCE or both) improved decoding accuracies considerably. Overall highest decoding accuracy was achieved by a the combination ReHo$||$fALFF$||$FCE and the highest number of vertices with statistical significant decoding accuracy by the combination ReHo$||$fALFF.

To illustrate the effect of combining rs-fMRI metrics more specifically across the cortex, figures \ref{fig:tspo_comparison} (D) and (E) illustrate differences in decoding accuracies and statistically significant areas between ReHo$||$fALFF$||$FCE and all other combinations. In comparison to when incorporating FCE only, a combination of all three rs-fMRI metrics (ReHo$||$fALFF$||$FCE) could considerably increase decoding accuracy across the cortex and yielded significant alterations in various typical resting-state networks \citep{Yeo2011_atlas}, as shown in figures \ref{fig:tspo_comparison} (D1) and (E1).
Due to the administration of alprazolam we found significant modulations in large parts of the visual resting-state network across the full occipital cortex (illustrated in red in figure \ref{fig:tspo_comparison} (E1)). In the default mode network we observed effects in the parietal cortex, inferior frontal gyrus, superior temporal gyrus, and the anterior and posterior cingulate cortex. In the dorsal attention resting-state network we found differences within the inferior temporal cortex, superior parietal lobule, angular gyrus and precentral gyrus. Further, we observed modulations predominantly in the frontal parts of the frontoparietal network, and modulations of the ventral attention network mainly in the anterior and posterior cingulate gyrus. In the somatomotor resting-state network we found an effect in the precentral gyrus. A detailed overlay of these comparisons with the resting-state atlas defined by \cite{Yeo2011_atlas} is provided in supplementary figure S5.
Figures \ref{fig:tspo_comparison} (D2) and (E2) show the comparison to the case when using only ReHo for decoding. Using ReHo also allowed us to predict alprazolam treatment in the visual, default and parts of the right dorsal attention and motor network (overlap highlighted in white), but not in any other area. A small number of vertices mainly within the left ventral attention network showed significant decoding accuracy when incorporating ReHo only but not when using the combination ReHo$||$fALFF$||$FCE (highlighted in blue). 
A comparison to fALFF in figures \ref{fig:tspo_comparison} (D3) and (E3) highlights that both variants were able to detect significant alterations within the visual network in both hemispheres, but using only fALFF did not allow us to decode in large parts of the right default, motor, frontoparietal and dorsal attention network. 
As shown in  figures \ref{fig:tspo_comparison} (D4) and (E4) adding FCE to ReHo made it possible to additionally predict the treatment in the superior temporal cortex, but reduced the prediction performance in the right frontal cortex. 
Finally, the comparison of ReHo$||$fALFF$||$FCE to the combinations  fALFF$||$FCE (figures \ref{fig:tspo_comparison} (D5) and (E5)) and ReHo$||$fALFF (figures \ref{fig:tspo_comparison} (D6) and (E6)) illustrates that combinations of fALFF with a connectivity based metric, either ReHo or FCE or both, yielded overall very similar predictions, whereby ReHo added some information on modulations in the right default mode network, and FCE on modulations in the superior temporal cortex.

In addition we compared the SVC with RBF kernel to alternative MVPA decoding models, and also studied the impact of different SL raidii in supplementary figures S6 (A) and (B). We found that also on this rs-fMRI dataset a SVC with RBF kernel yielded best results, and a SL radius of order 3 provided a reasonable trade-off between performance and computation time. A comparison to alternative variants of the three rs-fMRI metrics illustrates in supplementary figure S6 (C) that fALLF was slightly more informative than ALFF, FC efficiency was slightly more informative than FC betweeness centrality, and ReHo with a neighborhood order of 5 was more characteristic than of order 3 for the alprazolam treatment, whereby these differences were not statistically significant. The effect of removing the within subject mean of each rs-fMRI metric is depicted in supplementary figure S7. Thereby removing this baseline variability of rs-fMRI metrics across subjects has shown to considerably improve the prediction of pharmacological treatment.

\begin{figure}[htbp]\begin{center}\includegraphics[width=0.8\textwidth]{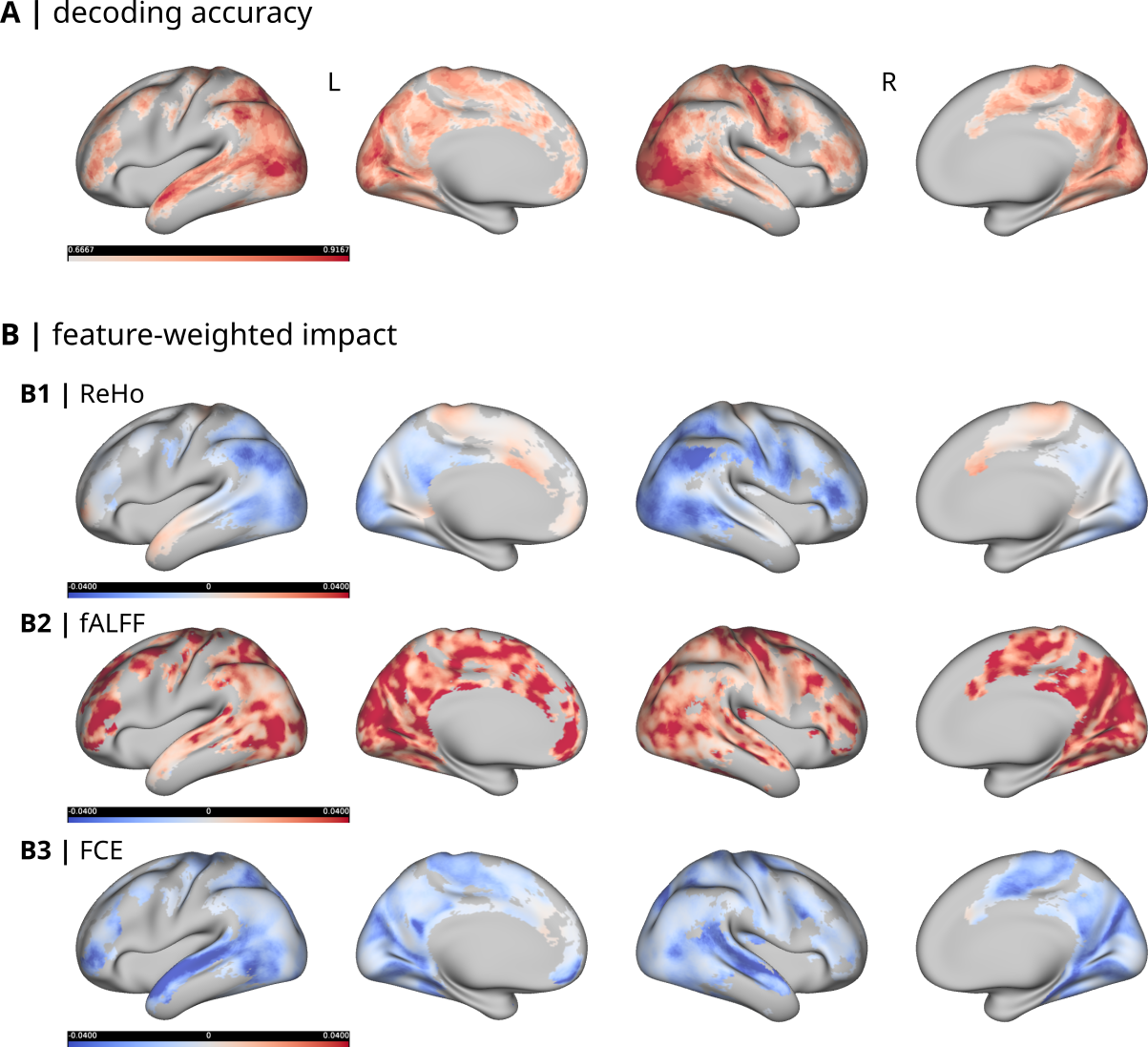}
\caption{(A) Combining ReHo, fALFF and FCE in FuSL yielded highest decoding accuracies in the visual, default mode and motor network. (B) The feature-weighted impact illustrates the informativeness, importance and relative increase or decrease of these metric, related to the treatment of alprazolam. (B1) Illustrates the decrease of ReHo mainly in the visual and frontoparietal network, including frontal parts of the default mode network. Simultaneously we observe an increase in ReHo in the somatomotor network, superior temporal cortex and anterior cingulate cortex. (B2) A characteristic increase in fALFF was found widespread in the visual, frontoparietal and default mode network. (B3) A high feature-weighted impact related to a decrease of FCE was highest within the visual network and default mode network within the superior temporal gyrus.}
\label{fig:tspo_shap}\end{center}\end{figure}

We then studied in detail the impact of each metric in the FuSL that combined all three rs-fMRI metrics (ReHo$||$fALFF$||$FCE). Figure \ref{fig:tspo_shap} (A) shows statistically significant decoding test accuracies across the cortex. 
Within these regions we observed highest discrimination capabilities in the visual network within the extrastriate cortex, in the default mode network within left angular gyrus, and within the right motor cortex. Additional performance metrics like decoding sensitivity, specificity and area under receiver operating characteristic (ROC) curve are visualized in supplementary figure S8. Analyzing the impact and feature-weighted impact of individual rs-fMRI metrics allowed us to relate the decoding performance to the individual input metrics. We therefore first compared two different variants for approximating Shapley values in supplementary figure S9, one based on permutation testing and one using linear explanation models \citep{Lundberg2017_shap}. Both approaches yielded similar impact maps across metrics, whereby the permutation based method was considerably more computationally efficient.
A comparison between the impact and feature-weighted impact shows that the magnitudes of these measures coincide qualitatively very well (supplementary figure S10), but analyzing the feature-weighted impact additionally allowed us to identify whether a metric increased or decreased under alprazolam treatment. In figure \ref{fig:tspo_shap} (B1) we observed that for ReHo a decrease within the visual network was indicative of the administration of alprazolam. Further, we observed a characteristic decrease in ReHo mainly in the frontoparietal resting-state network, and within the default mode network a decrease in the parietal and posterior cingulate cortex. Simultaneously we found an increase in ReHo in the somatomotor network, and within the default mode network an increase in the anterior cingulate cortex and superior temporal cortex. Compared to ReHo, for fALFF we generally observed higher feature-weighted impact values across the cortex, as illustrated in figure \ref{fig:tspo_shap} (B2). A characteristic increase in fALLF due to alprazolam was found bilaterally within the visual network in the extrastriate cortex. Additionally, we found an increase with high impact in fALLF within the frontoparietal and default mode network, in frontal cortex, and anterior and posterior cingulate cortex. In relation to alprazolam we found predominantly a decrease in FCE as shown in figure \ref{fig:tspo_shap} (B3). This decrease in FCE  had highest impact on decoding also within the visual network in the extrastriate cortex, and within the default mode network in the superior temporal gyrus. Additionally, an overlay of the decoding accuracy map and feature-weighted impact maps with the resting-state network atlas established by \cite{Yeo2011_atlas} is provided in supplementary figure S11.

\section{Discussion}

In our study we presented an interpretable SL framework for fusing the complementary information of different rs-fMRI metrics. We first used an artificial dataset to investigate how different combinations of informative and uninformative data sources can affect decoding capabilities of FuSL. We demonstrated that across a variety of different artificial datasets the gain of adding informative sources exceeded the loss in performance when adding an uninformative source. We further illustrated how reconstructing feature-importance using SHAP allowed us to recover the impact of each data source, as well as the exact location of informative vertices. In a next step we studied how combining ReHo, fALFF and FCE can enhance the prediction of pharmacological treatment in rs-fMRI. 
In this application combining fALFF with connectivity-based metrics (ReHo, FCE or both) could considerably enhance the decoding performance.
Especially in clinical studies where collecting a very large sample size is often not feasible, the integration of different information sources can help improving decoding sensitivity on smaller datasets. Still, how distinct rs-fMRI markers are coupled and related to each other is not yet investigated and understood in detail \citep{Lv2018_review_rs-fMRI}. Our comparisons show that metrics derived from the same signal can exhibit very distinct characteristics in different areas across the cortex. For instance in the somatomotor network we observed an increase both in fALFF and ReHo, but within the visual network we observed an increase in fALFF whereas ReHo decreased. Note that the interrelations of rs-fMRI markers and their optimal combination in FuSL might therefore also vary between different rs-fMRI studies. It has yet to be shown which of the patterns reported here will remain stable across different studies.

In a next step we have shown how one can utilize explainable AI to reconstruct the impact of each rs-fMRI metric on the decoding in our workflow. Incorporating the SHAP framework \citep{Lundberg2017_shap} allowed us to analyze locally the importance of features (in our study rs-fMRI metrics), which we combined with local informativeness (derived from decoding accuracy) to quantify the impact of a feature across the cortex. We propose to additionally incorporate local differences in features between groups to visualize whether a feature-expression increases or decreases. In theory a large feature difference does not necessarily correlate with a high feature impact, because the output of a non-linear decoding model can depend on a highly non-linear combination of its input features. In our practical application we however observed that absolute amplitudes of the impact and feature-weighted impact corresponded qualitatively very well (supplementary figure S10), so for a compact interpretation we considered it useful to visualize the feature-weighted impact. Yet, to study the impact of a feature in more detail, we propose to additionally visualize the impact in isolation. Besides using SHAP for computing the differential impact of each metric, we have shown that incorporating explainable AI can be generally beneficial for searchlight analysis. Evaluating feature-importance of the decoding model could help to reconstruct the exact location of informative vertices or voxels, which makes this technique interesting for all types of searchlight decoding analysis.

Studying the feature-weighted impact of individual rs-fMRI markers could reveal different effects of the alprazolam treatment on brain functions. For example, the sedation-typical increase in fALFF \citep{Kiviniemi2003, Kiviniemi2005} and decrease in FCE \citep{Wei2013_connectivity_ch, Hashmi2017_dexm_sedation, Schrouff2011_propofol_wb_conn} in low-level sensory regions could relate to several well-studied behavioral side effects of alprazolam. Observed modulations in the motor and visual cortices might be connected to motor coordination impairments and visuospatial and visuomotor abilities related to benzodiazepines \citep{Griffin2013_benzos_motor, Golombok1988_cognitive_benzos, Tata1994_cogn_recovery_benzos}. The decrease in ReHo and FCE in more high-level areas might be connected to impairments in cognitive domains such as attention, working memory and semantic processing, which are affected by long-term treatments with benzodiazepines \citep{Barker2004_benzos_cognition, Stewart2005_benzos_cognition}.

However, at this point the interpretation of above inconsistent directional changes of ReHo in different brain areas need to be regarded with some caution and seeks deeper investigations. Furthermore, the detailed mechanistic connections between pharmacological interventions and modulations in BOLD dynamics are relatively poorly understood to date, so also inferences from alterations in rs-fMRI markers to behavioral changes should consider these current knowledge gaps \citep{Mahani2017_review_rs_pharma}.
Within the scope of our study we investigated the application of FuSL to rs-fMRI using only one dataset, so finding reliable and optimal combinations of rs-fMRI metrics likely requires more extensive research including potentially larger datasets. Furthermore, to what extent combining information from very distinct neuroimaging modalities improves decoding capabilities of FuSL has to be shown in future research.
Until now our study could provide first evidence for the potential of FuSL in MVPA decoding applications. While deep learning frameworks often require a large amount of training data and manual fine-tuning of model parameters \citep{Schulz2020_scaling}, SL has proven to be a robust and widely-used MVPA tool that limits its complexity inherently by using spatially localized information.
Furthermore, by using a model-agnostic approach for measuring feature importance based on Shapley values \citep{Lundberg2017_shap}, our workflow allows one to integrate any kind of MVPA decoding model. Besides this application of decoding in rs-fMRI, our FuSL workflow can be easily adapted to various additional domains of neuroimaging. Combining different imaging modalities in FuSL could be beneficial for studying processes and conditions that simultaneously affect brain structure and function like observed in learning \citep{Carreiras2009_learning} or schizophrenia \citep{Dabiri2022_schizo}. Also this framework would allow one to integrate activation maps of combined fMRI and EGG studies \citep{Menon2005_eeg_fmri}. In this manner FuSL provides a flexible and interpretable tool for diverse applications in neuroimaging analysis.


\section*{Data and Code Availability}

The codes including the artificial dataset are publicly available at: \url{https://github.com/simonvino/FuSL}.
The resting-state fMRI dataset is available from the corresponding author on reasonable request.

\section*{Author Contributions}

SW and JS designed the analysis. SW implemented the analysis. RR and CN were responsible for the clinical MRI study. MR, LMB and JS acquired the imaging data. SW, and JS wrote the initial draft. All authors revised the manuscript.

\section*{Funding}

This work has been supported by the German Research Foundation (Deutsche Forschungsgemeinschaft) (DFG), project number 403161218, to JS within the framework of FOR2858. 

\section*{Declaration of Competing Interests}

The authors declare no conflicts of interest.


\clearpage

\bibliography{manuscript_arxiv}


\clearpage

\setcounter{figure}{0}
\renewcommand{\thefigure}{S\arabic{figure}}
\renewcommand{\thetable}{S\arabic{table}}
\renewcommand{\theHfigure}{Supplement.\thefigure}

\section*{Supplementary Information}

\subsection*{Supplement I: Definitions}

\subsubsection*{Resting-state fMRI metrics}

\paragraph{Regional homogeneity}

We first investigate local connectivity, as defined by \textit{regional homogeneity} (ReHo). Thereby ReHo characterizes the coherence of the BOLD signal within a local neighbourhood of voxels or vertices \citep{Zang2004}. In a neighbourhood containing $ N $ vertices, we computed ReHo as the average Pearson correlation coefficient $ r_{ij} $ between all pairs of timecourses $ i $ and $ j $: 

\beq
    ReHo = \frac{\sum_{i \not= j} r_{ij}}{N(N-1)}
\eeq

\paragraph{Low frequency fluctuations}

To further analyze changes in spectral characteristics of the BOLD signal, we study alterations in \textit{fractional amplitude of low frequency fluctuation} (fALFF) values across the cortex \citep{Zou2008}. For each vertex $ i $ fALFF can be computed as the ratio of the power of the BOLD signal $ S_i(t) $, after being filtered with a bandpass filter $ h(t) $, to the power of the unfiltered signal $ S_i(t) $:

\beq
    fALFF = \sqrt{\frac{\sum_{t} (h(t) * S_i(t))^2}{\sum_{t} S_i(t)^2}}
\eeq
Here $ * $ denotes the convolution operation and $ t $ the temporal index. In our application we focused on the very low frequency range $ 0.01Hz-0.05Hz $, which has shown to be characteristic for sedation effects observed in resting-state fMRI \citep{Kiviniemi2000, Kiviniemi2003, Kiviniemi2005}.

\paragraph{Functional connectivity efficiency}

\textit{Functional connectivity} (FC) between two brain regions $ i $ and $ j $ was computed as the Pearson correlation coefficient $ r_{ij} $ between the averaged BOLD signals of these regions. When computing these correlation values between all $ N $ regions, a FC network can be characterized by an adjacency matrix $ \mb{A} \in \mbb{R}^{N \times N} $, whereby one entry $ a_{ij} $ of this matrix describes the FC strength between brain region $ i $ and $ j $. The \textit{shortest path length} $ d_{ij} $ between node $ i $ and node $ j $ in a network can be defined as the minimum number of edges traversed in an optimal path between those nodes. Based on this definition, the connection \textit{efficiency} $ E_{ij} $ between two nodes $ i $ and $ j $ can be derived \citep{Sporns2010}:

\beq
    E_{ij} = \frac{1}{d_{ij}}
\eeq
The average FC efficiency (FCE) of a ROI $ i $ is thereby defined as:

\beq
    E_{i} = \sum_{j} E_{ij}
\eeq
The centrality of a node $ i $ in a graph $ G $ can be described by \textit{betweenness centrality}, defined as the sum of the fraction of shortest paths between two nodes $ j $ and $ k $ that pass through node $ i $:

\beq
    c_B(i) = \sum_{j,k \in G} \frac{\sigma(j,k | i)}{\sigma(j,k)}
\eeq
where $ \sigma(j,k) $ is the number of shortest paths between $ j $ and $ k $, and $ \sigma(j,k | i) $ the number of shortest paths passing through node $ i $ (other than $ j, k $).

\subsubsection*{Searchlight feature importance}

To recover the importance of each input metric for decoding a specific brain state we used \textit{Shapley additive explanations} (SHAP) \citep{Lundberg2017_shap}. SHAP is an additive feature attribution method that allows to relate the contribution of input features to the prediction of a model. For a single input $ x $ the predictions of a model $ f(x) $ are approximated by a linear relationship:

\beq
    g(z') = \phi_0 + \sum_{i=1}^M \phi_i z_i'
\eeq
with the explanation model $ g(z') $ and coalition vector $ z' \in \{0, 1\}^M$, indicating whether a feature is present or not, $M$ the number of input features and Shapley values $\phi_i \in \mbb{R} $ of feature $ i $. 
These values can be approximated using different model specific and model agnostic methods described by 
\cite{Lundberg2017_shap}, whereby in our implementation we rely on the model-agnostic approach employing successive permutations of input features. We additionally compare this permutation based method for approximating Shapley values to linear explanation models (KernelSHAP). We then compute the importance of an input feature $ i $ as the average of absolute Shapley values of all $ J $ test samples:

\beq
    I_i = \frac{1}{J} \sum_{j=1}^J | \phi_i^{(j)} |
\eeq
In our FuSL framework the input features of one SL decoder reflect the different rs-fMRI metrics $ m $ at different vertex locations $ v $ on the cortex. For each individual SL $ k $ we get in this manner an importance score $ I_{mv}^{(k)} $, representing the contribution of metric $ m $ at vertex $ v $ in SL $ k $. To obtain spatial maps that reflect the importance of a metric $ m $ across vertices $ v $, we average over the importance scores of all SLs $k$ that overlap in vertex $ v $:

\beq
    I_{mv} = \frac{1}{K} \sum_{k=1}^K I_{mv}^{(k)}
\eeq
In a next step, to combine feature importance with the actual information content about a brain state at a specific location, we weight the importance $I_{mv}$ with the decoding accuracy score $ s_v $ at each vertex $ v $ to compute the impact score defined as:

\beq
    \Psi_{mv} = I_{mv} \cdot s_v
\eeq
To additionally visualize, whether the values $ X_{mv}^{(j)} $ of a metric $ m $ at vertex $ v $ are relatively increased or decreased between groups or treatments, we computed the average difference between samples $ p $ of group $ G_1 $ and samples $ q $ of group $ G_2 $:

\beq
     \Delta_{mv} = \frac{1}{P} \sum_{p \in G_1 }^{P} X_{mv}^{(p)} - \frac{1}{Q} \sum_{q \in G_2 }^{Q} X_{mv}^{(q)} 
\eeq
Based on these differences we derived the feature-weighted impact of metric $ m $ at vertex $ v $ as:

\beq
     \Phi_{mv} =  \Psi_{mv} \cdot \Delta_{mv} 
\eeq

\vspace{1cm}
\clearpage


\subsection*{Supplement II: Supplementary figures and tables} \label{sec:fig_tabs}

\begin{figure}[!htb]
\bc
\makebox[\textwidth][c]
{
\includegraphics[width=0.99\textwidth]{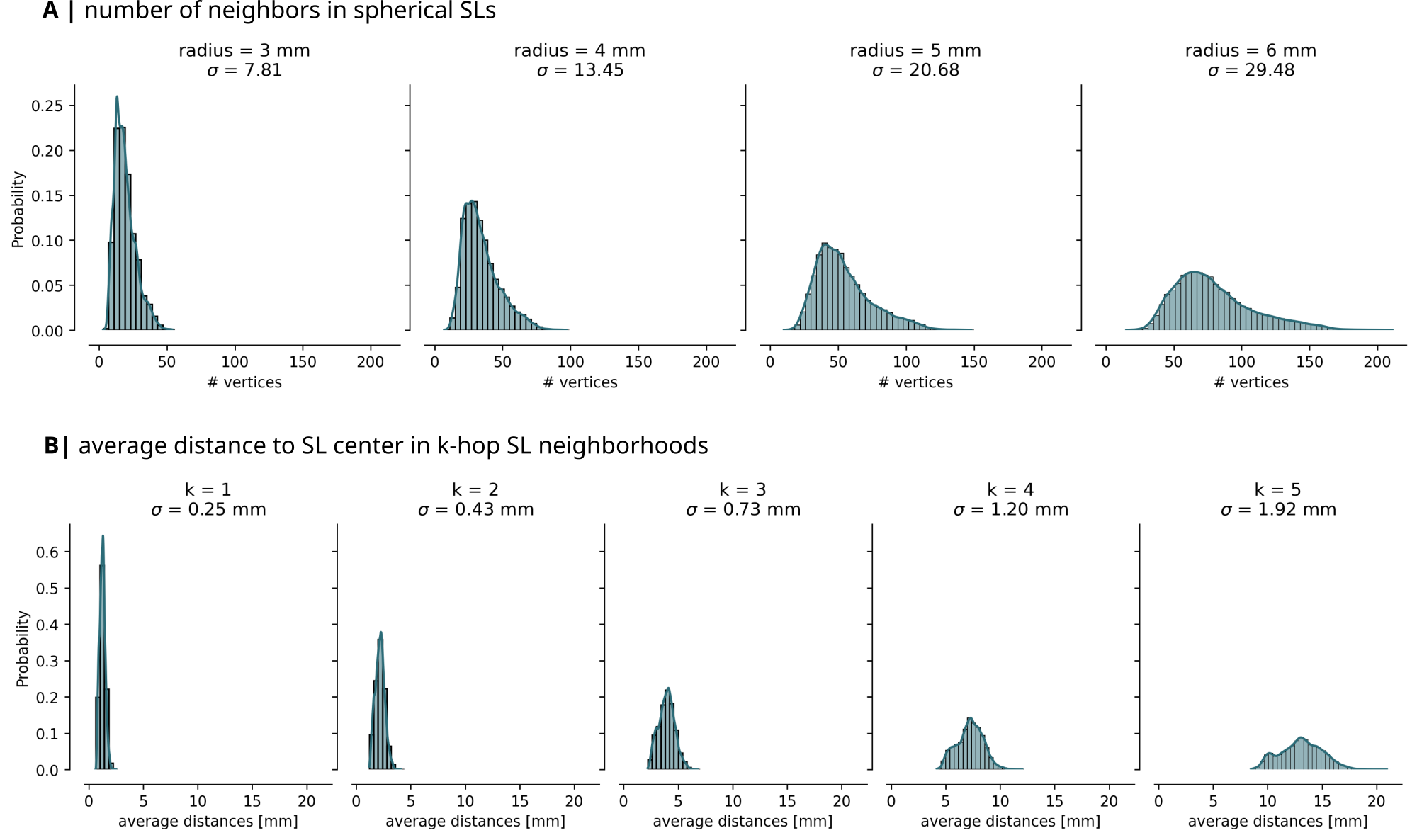} 
}
\ec 
\caption{Spatial variability in the mesh density of the fsLR standard space mesh. (A) Distribution of number of vertices in a spherical SLs defined in the worldspace across the cortex. A larger radius (in $mm$) increases the standard deviation of number of vertices $ \sigma $ within spherical SLs. (B) Average distance of vertices to SL center when using a constant number of k-hop neighbors. The variability in the size of the SLs across the cortex increases with a larger number of k-hop neighbors included in SLs.}
\end{figure}


\begin{figure}[!htb]
\bc
\makebox[\textwidth][c]
{
\includegraphics[width=0.8\textwidth]{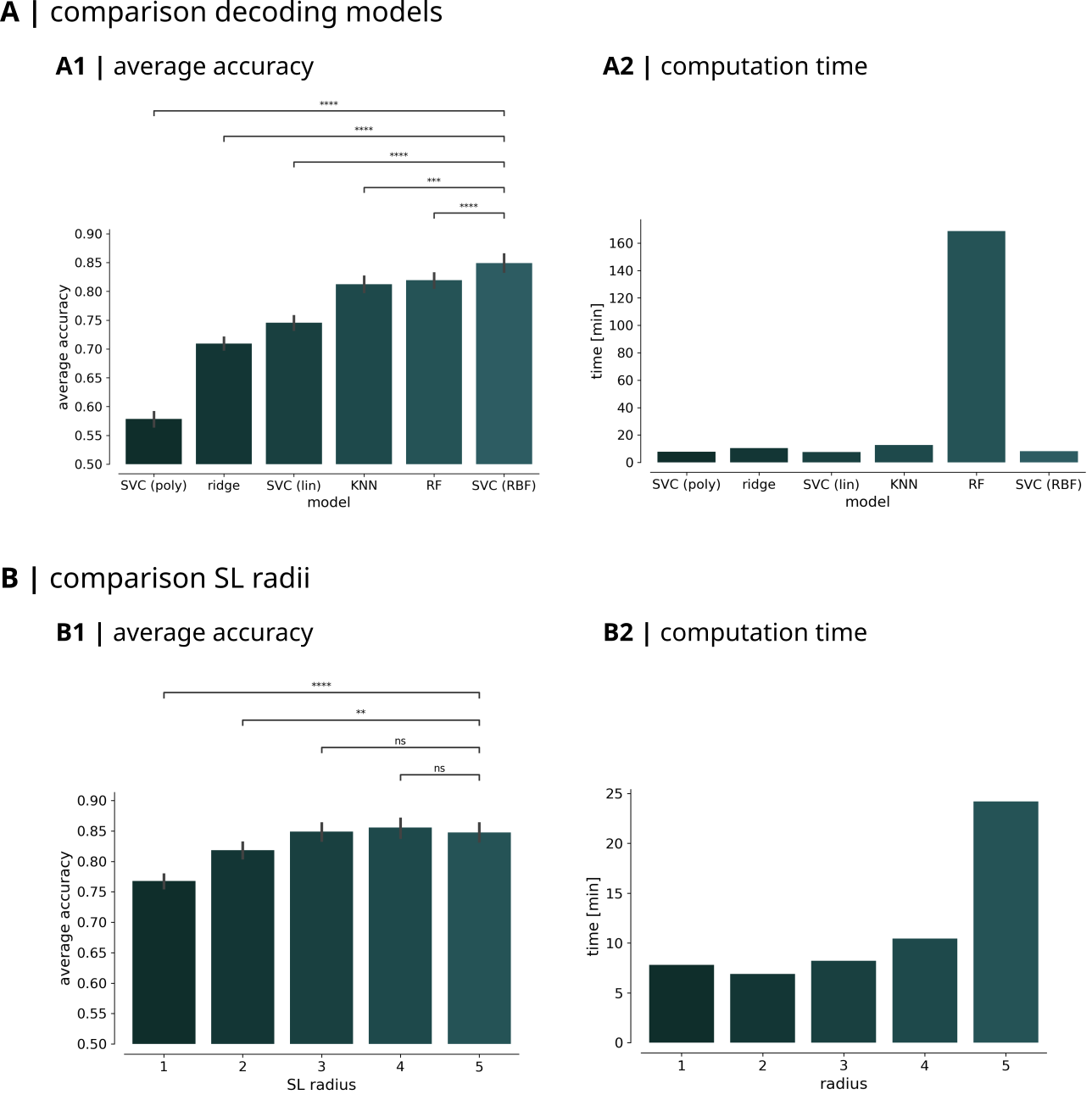} 
}
\ec 
\caption{Impact of MVPA model and search light (SL) radius on the decoding using artificial data. (A) A support vector classifier (SVC) with radial basis function kernel (RBF) outperforms SVC with a linear (lin) and polynomial (poly) kernel. Further it shows better decoding accuracy than a k-nearest neighbor (KNN) classifier, a random forest (RF) classifier based on 50 decision trees, and a ridge regression classifier. (B) A SL radius of order 3 provides a reasonable trade-off between accuracy and computation time. Error bars represent 95\% confidence intervals across folds. Significant differences of accuracies are indicated with: *: $ p \leq 0.05 $, **: $ p \leq 0.01 $, ***: $ p \leq 0.001 $, ****: $ p \leq 0.0001 $, ns: not significant, (ns): not significant after false discovery rate correction.}
\end{figure}


\begin{figure}[!htb]
\bc
\makebox[\textwidth][c]
{
\includegraphics[width=0.95\textwidth]{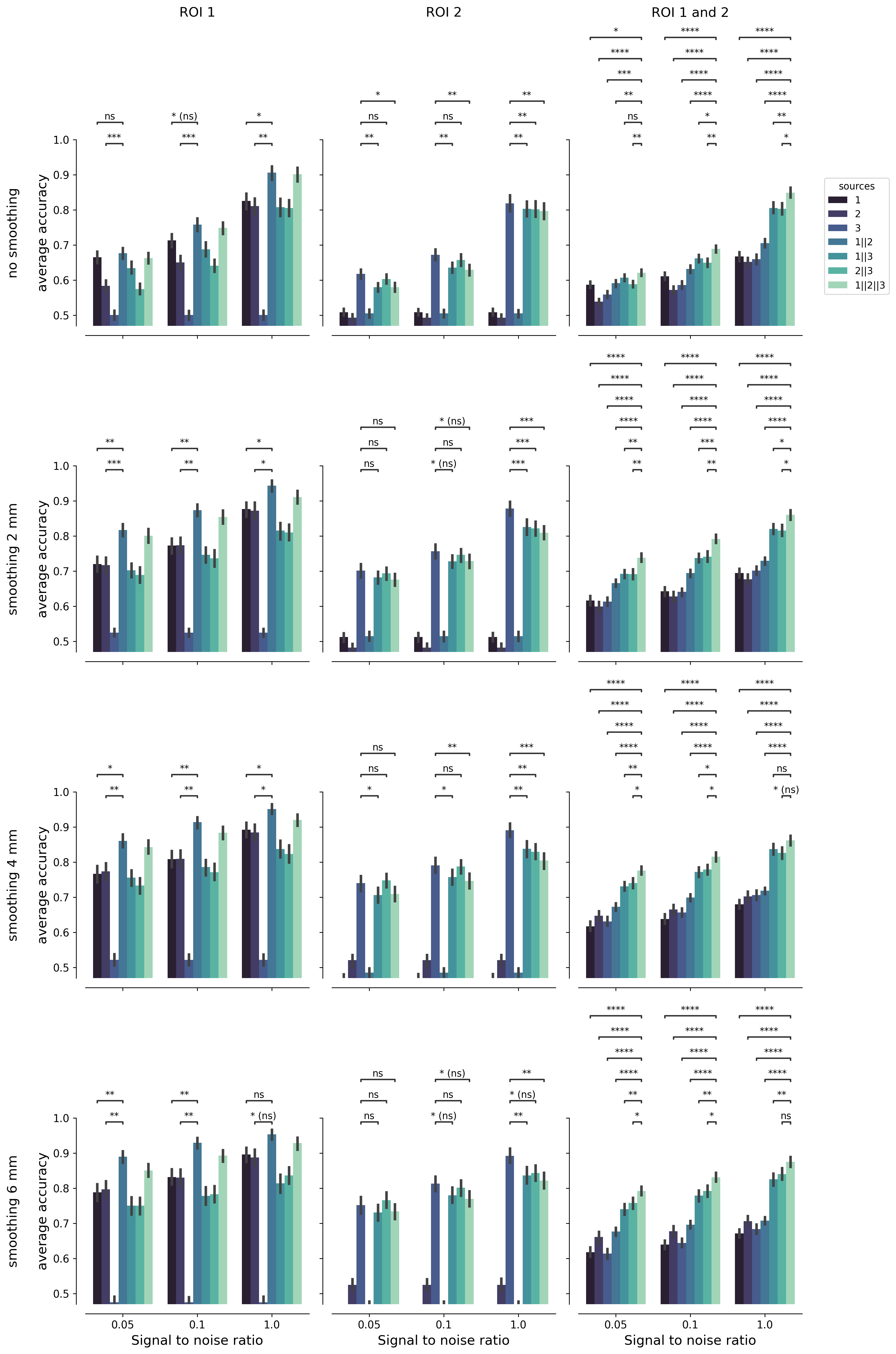} 
}
\ec 
\caption{Decoding performance of FuSL on an artificial dataset with a signal standard deviation of $\sigma = 1 $ across samples, in dependence of different signal to noise ratios and different levels of spatial autocorrelations simulated by Gaussian smoothing. Error bars represent 95\% confidence intervals across folds. Significant differences of accuracies are indicated with: *: $ p \leq 0.05 $, **: $ p \leq 0.01 $, ***: $ p \leq 0.001 $, ****: $ p \leq 0.0001 $, ns: not significant, (ns): not significant after false discovery rate correction.}
\end{figure}


\begin{figure}[!htb]
\bc
\makebox[\textwidth][c]
{
\includegraphics[width=0.95\textwidth]{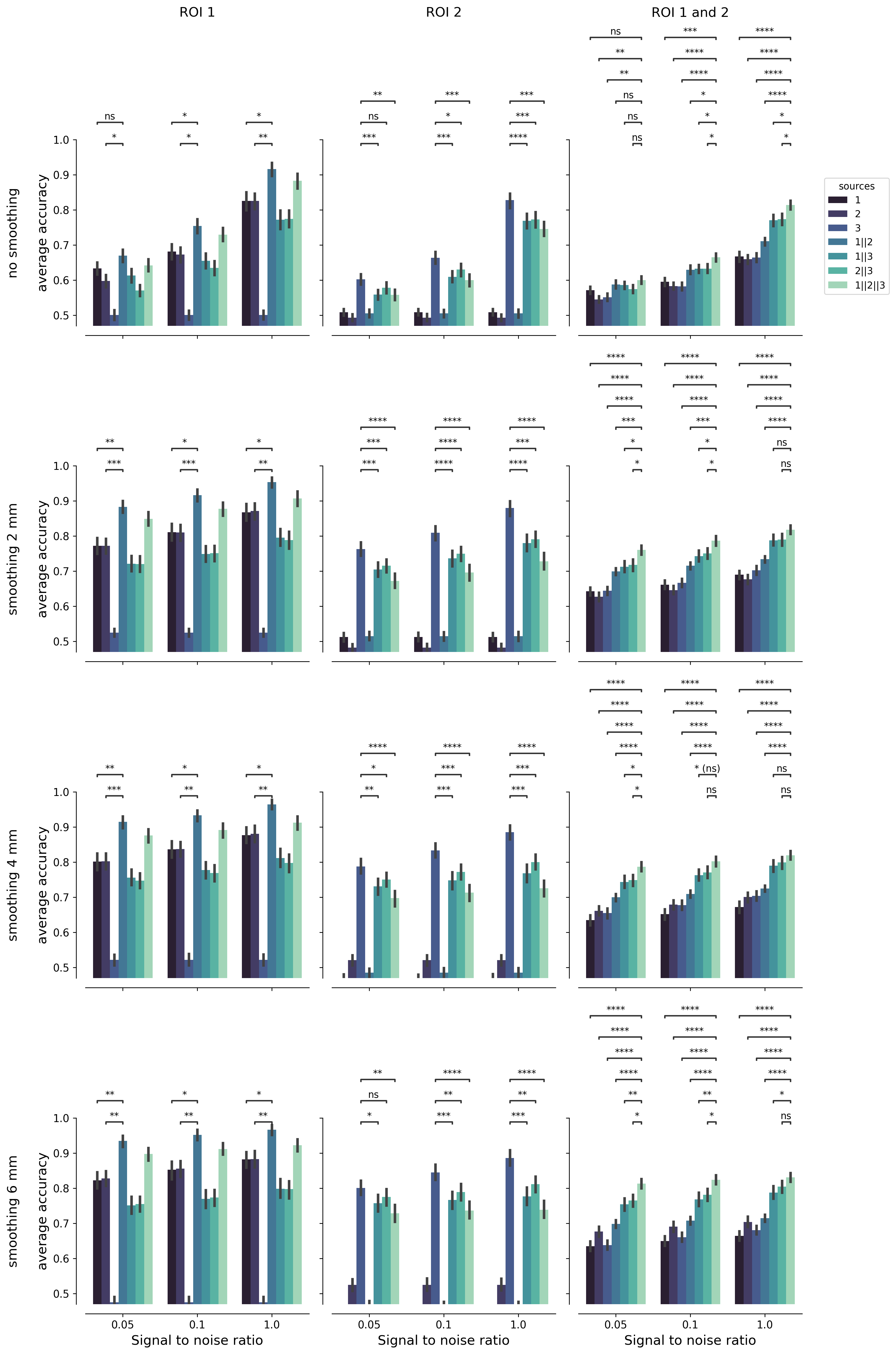} 
}
\ec 
\caption{Decoding performance of FuSL on an artificial dataset with a signal standard deviation of $\sigma = 2 $ across samples, in dependence of different signal to noise ratios and different levels of spatial autocorrelations simulated by Gaussian smoothing. Error bars represent 95\% confidence intervals across folds. Significant differences of accuracies are indicated with: *: $ p \leq 0.05 $, **: $ p \leq 0.01 $, ***: $ p \leq 0.001 $, ****: $ p \leq 0.0001 $, ns: not significant, (ns): not significant after false discovery rate correction.}
\end{figure}




\begin{table}[h!]
\centering
\caption{P-values and effect sizes (Cohen's d) related to the comparison of decoding accuracies between ReHo$||$fALFF$||$FCE and all other respective combinations. All differences that remained significant after correcting for multiple comparisons are marked in bold font.}
      \begin{tabular}{cccc}
        \hline
Comparison   & p-value & effect size \\ \hline
ReHo$||$fALFF$||$FCE vs. FCE & \textbf{0.0053} & 0.817 \\
ReHo$||$fALFF$||$FCE vs. ReHo & \textbf{0.0004} & 1.015 \\
ReHo$||$fALFF$||$FCE vs. fALFF & \textbf{0.0256} & 0.659 \\
ReHo$||$fALFF$||$FCE vs. ReHo$||$FCE & \textbf{0.0127} & 0.634 \\
ReHo$||$fALFF$||$FCE vs. fALFF$||$FCE & \textbf{0.0091} & 0.145 \\
ReHo$||$fALFF$||$FCE vs. ReHo$||$fALFF & 0.1998 & 0.107 \\
      \end{tabular}
\end{table}


\begin{figure}[!htb]
\bc
\makebox[\textwidth][c]
{
\includegraphics[width=0.99\textwidth]{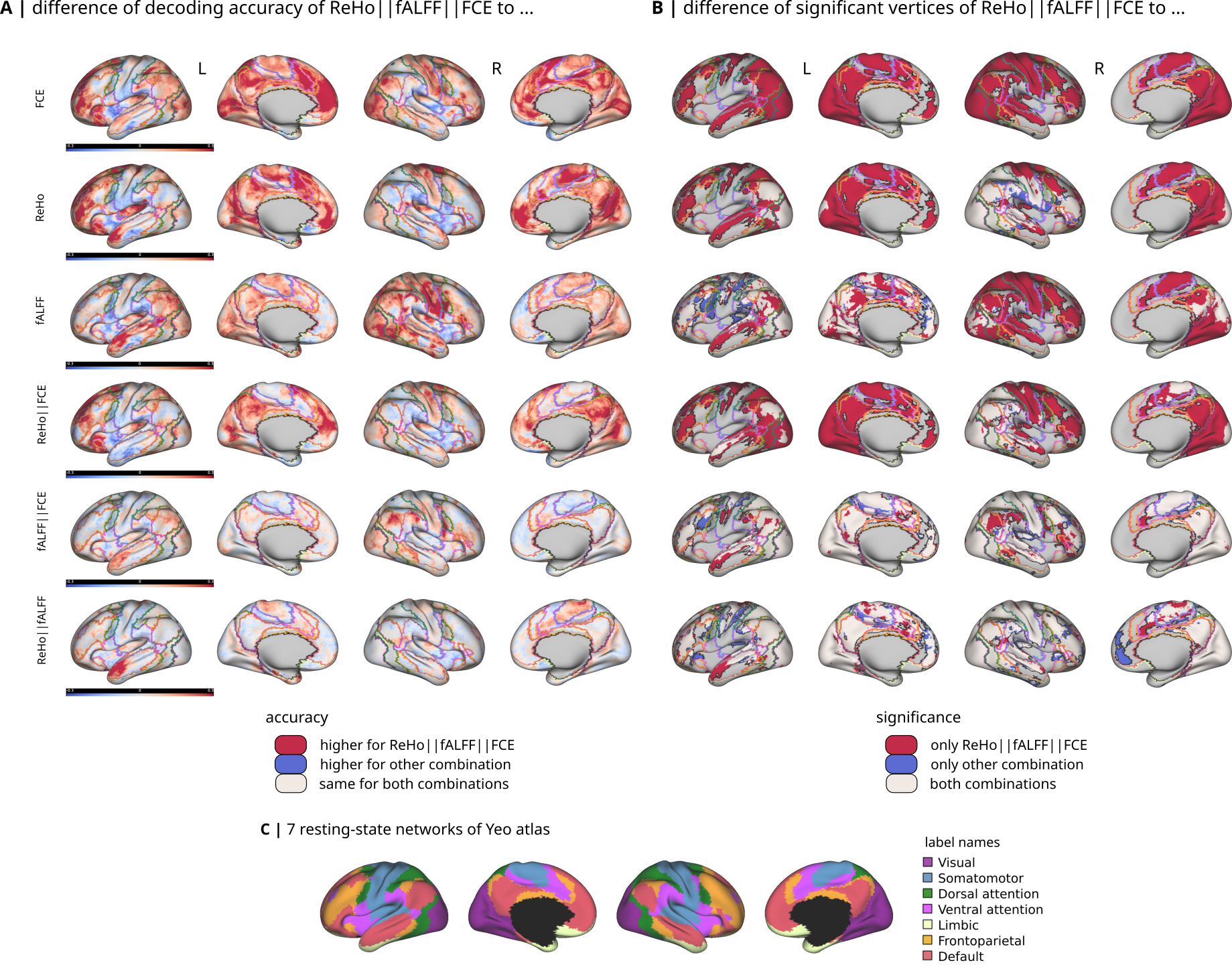} 
}
\ec 
\caption{A comparison of decoding accuracy and significant vertices of combination ReHo$||$fALFF$||$FCE and all other combinations overlayed with resting-state networks defined by \cite{Yeo2011_atlas}. (A) Local differences in decoding test accuracies between ReHo$||$fALFF$||$FCE and all other combinations in the left (L) and right (R) cortex. (B) Locations where only ReHo$||$fALFF$||$FCE yields significant decoding accuracies are highlighted in red. Locations where both combinations can significantly decode are marked in white, and locations where only the other combination is able to decode are depicted in blue. (C) Displays the 7 resting-state networks defined by \cite{Yeo2011_atlas}.}
\end{figure}


\begin{figure}[!htb]
\bc
\makebox[\textwidth][c]
{
\includegraphics[width=0.8\textwidth]{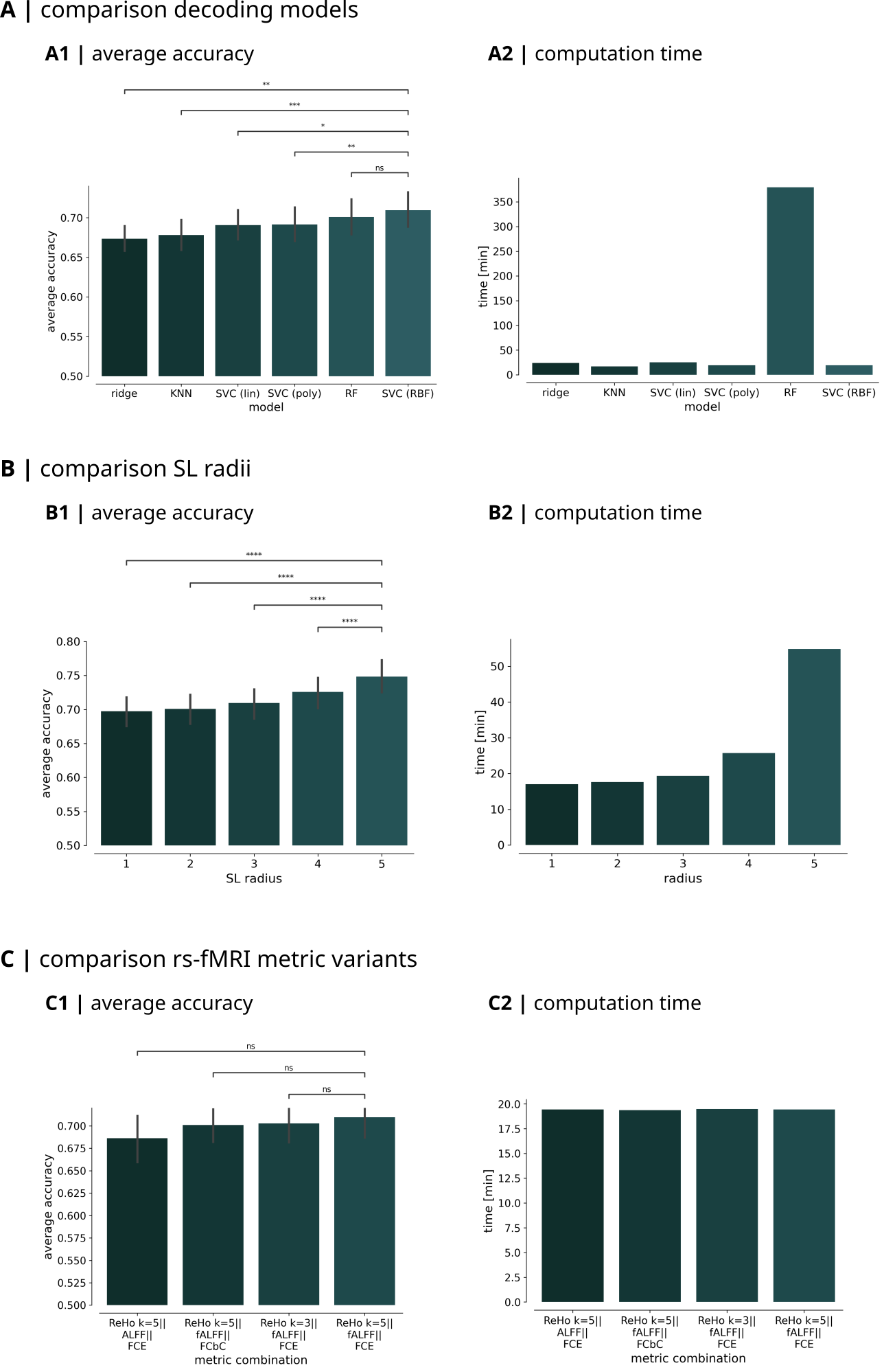} 
}
\ec 
\caption{Impact of MVPA model, search light (SL) radius and resting-state fMRI metric on the decoding of pharmacological treatment. (A) A support vector classifier (SVC) with radial basis function kernel (RBF) outperforms SVC with a linear (lin) and polynomial (poly) kernel. Further it shows better decoding accuracy than a k-nearest neighbor (KNN) classifier, a random forest (RF) classifier based on 50 decision trees, and a ridge regression classifier. (B) A SL radius of order 3 shows a reasonable trade-off between accuracy and computation time. (C) When combining different rs-fMRI metrics, fALLF shows to be slightly more informative than ALFF, FC efficiency (FCE) slightly more informative than FC betweeness centrality  (FCbC), and ReHo with a k-hop neighborhood order of 5 is slightly more informative than of order 3. Error bars represent 95\% confidence intervals across folds. Significant differences of accuracies are indicated with: *: $ p \leq 0.05 $, **: $ p \leq 0.01 $, ***: $ p \leq 0.001 $, ****: $ p \leq 0.0001 $, ns: not significant, (ns): not significant after false discovery rate correction.}
\end{figure}


\begin{figure}[!htb]
\bc
\makebox[\textwidth][c]
{
\includegraphics[width=0.6\textwidth]{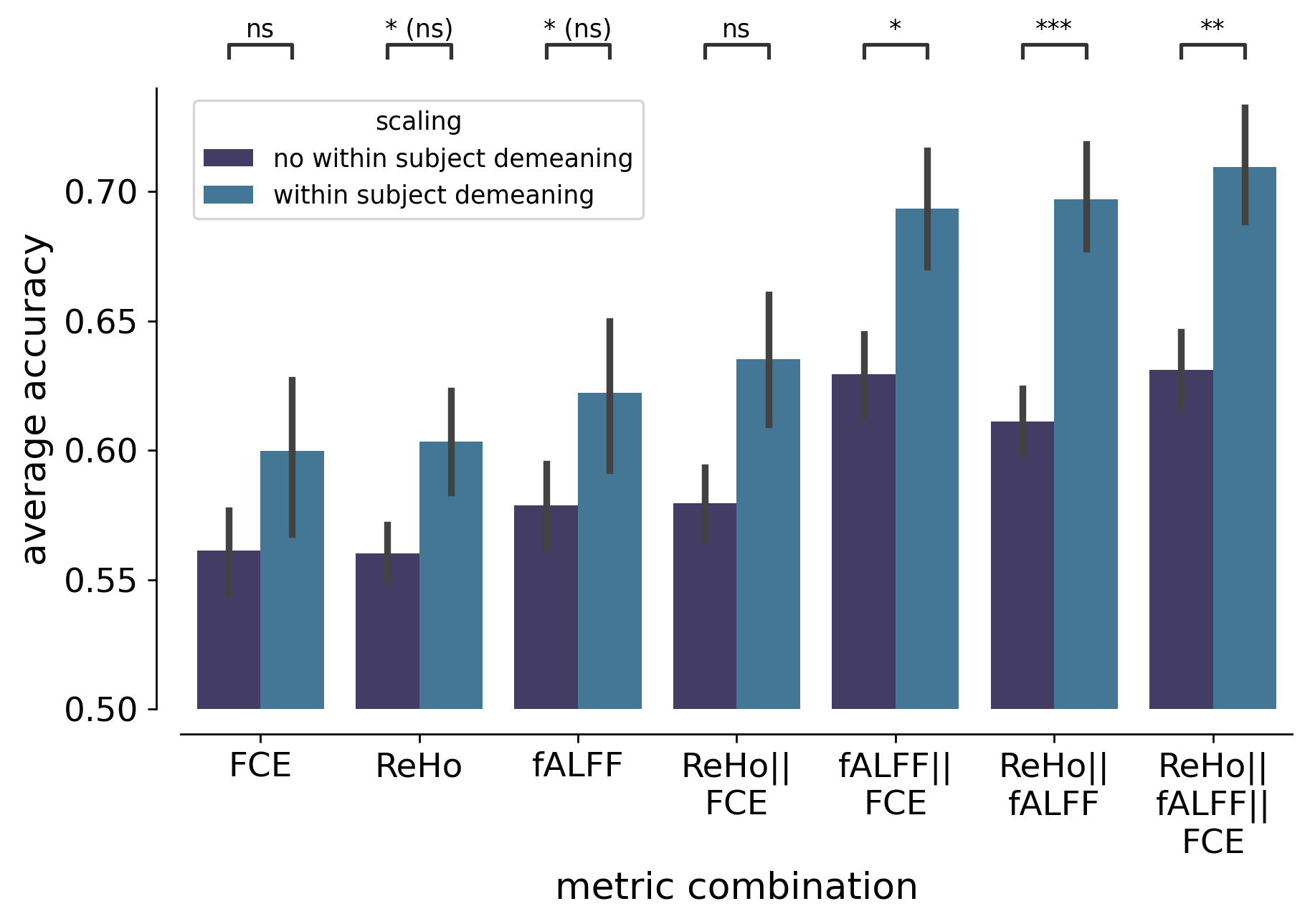} 
}
\ec 
\caption{Impact of within subject demeaning on the decoding accuracy. Removing the mean of rs-fMRI metrics within each subject removes the baseline variability of rs-fMRI metrics between different subjects and helps to better detect the effect of the alprazolam treatment, resulting inconsiderably higher decoding accuracies. Error bars represent 95\% confidence intervals across folds. Significant differences of accuracies are indicated with: *: $ p \leq 0.05 $, **: $ p \leq 0.01 $, ***: $ p \leq 0.001 $, ns: not significant, (ns): not significant after false discovery rate correction.}
\end{figure}
  

\begin{figure}[!htb]
\bc
\makebox[\textwidth][c]
{
\includegraphics[width=0.85\textwidth]{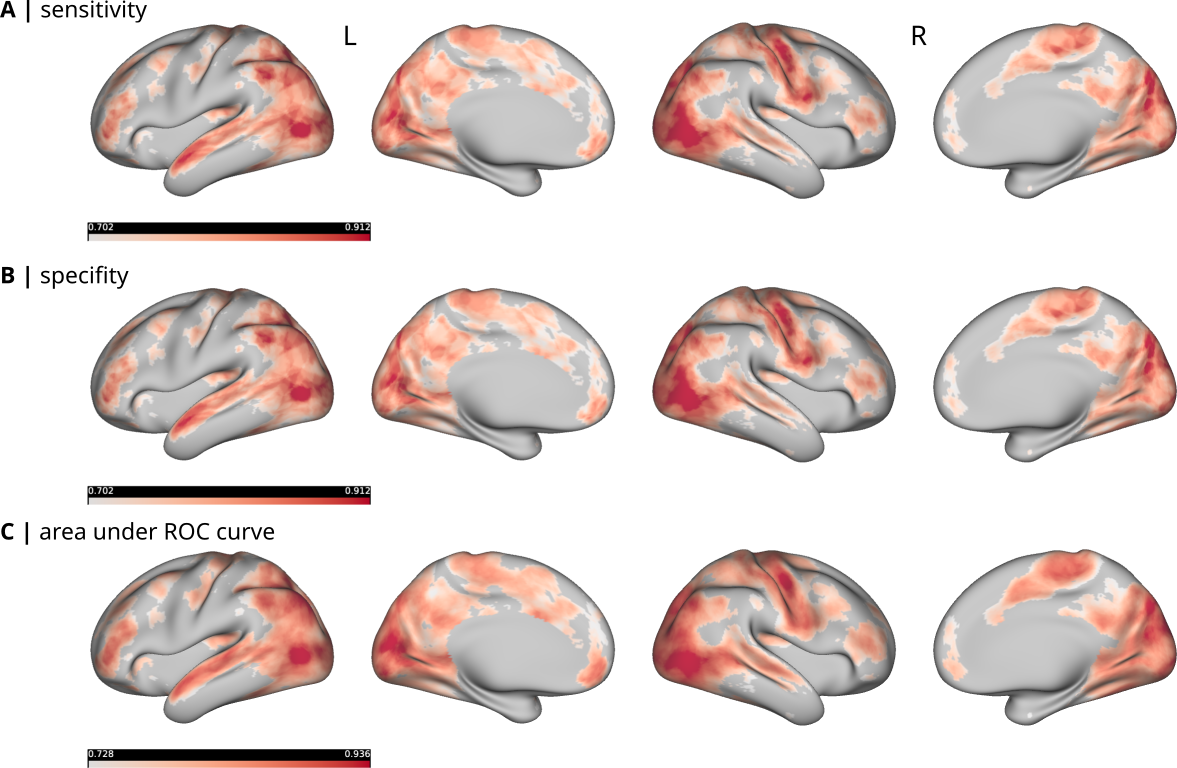} 
}
\ec 
\caption{Sensitivity, specificity and area under receiver operating characteristic (ROC) curve based on the decoding of the alprazolam treatment. 
Patterns of sensitivity (A), specificity (B) and area under ROC curve (C) resemble closely the pattern of decoding accuracy.
}
\end{figure}


\begin{figure}[!htb]
\bc
\makebox[\textwidth][c]
{
\includegraphics[width=0.85\textwidth]{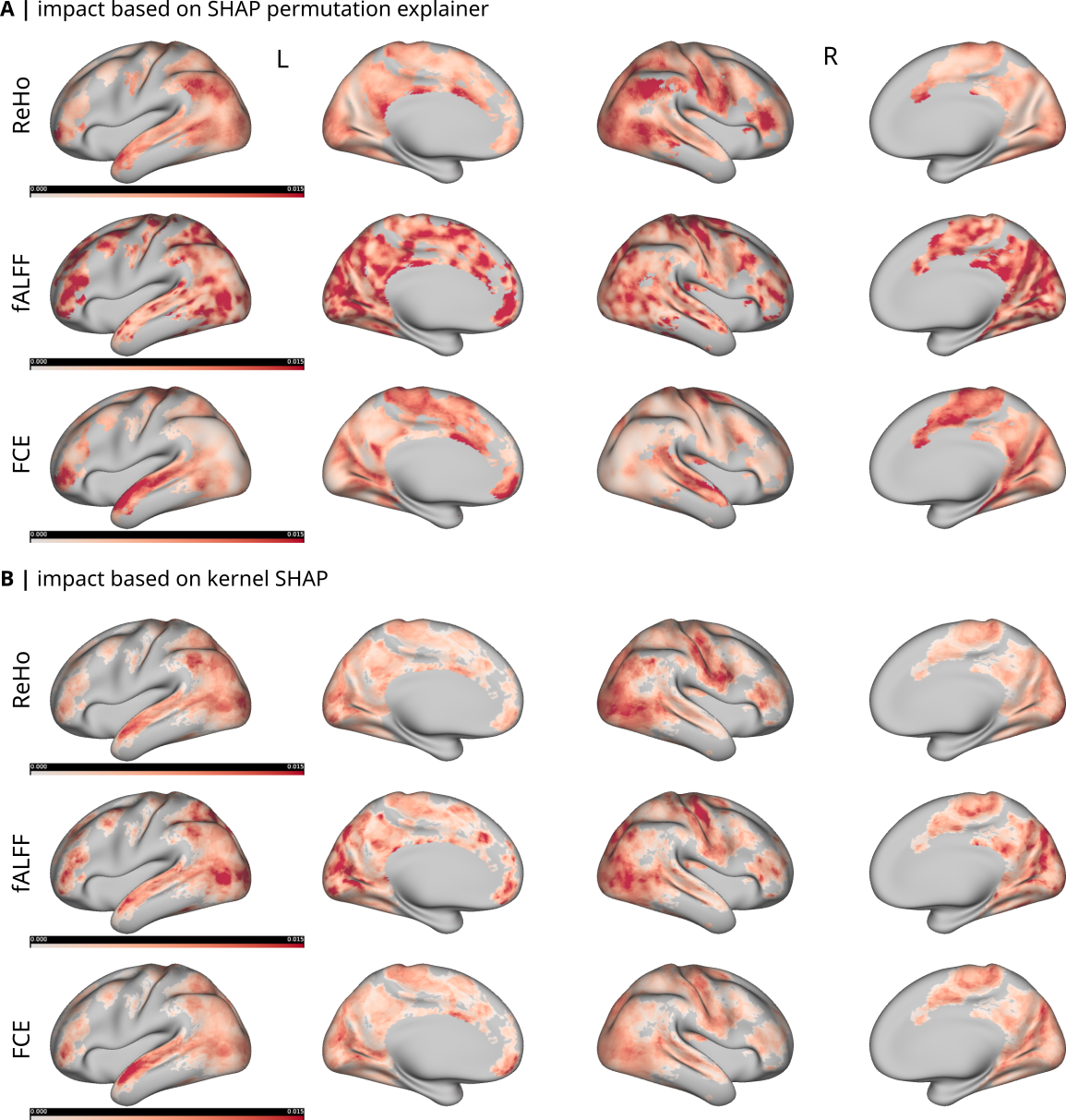} 
}
\ec 
\caption{Comparison of methods to compute SHAP values for determining the impact of each rs-fMRI metric. (A) Impact based on SHAP values derived from permutation testing. (B) Impact based on SHAP values derived from Kernel SHAP. Both explanation approaches produce similar impact maps, whereby Kernel SHAP attributes slightly less importance to fALFF, than the permutation based approach. The computation time of the permutation based method is with 5.29 h lower than the time of the Kernel SHAP method requiring 17.97 h on a Intel(R) Xeon(R) Platinum 8268 CPU, running under Ubuntu 20.04 with 256 GB of RAM.}
\end{figure}


\begin{figure}[!htb]
\bc
\makebox[\textwidth][c]
{
\includegraphics[width=0.85\textwidth]{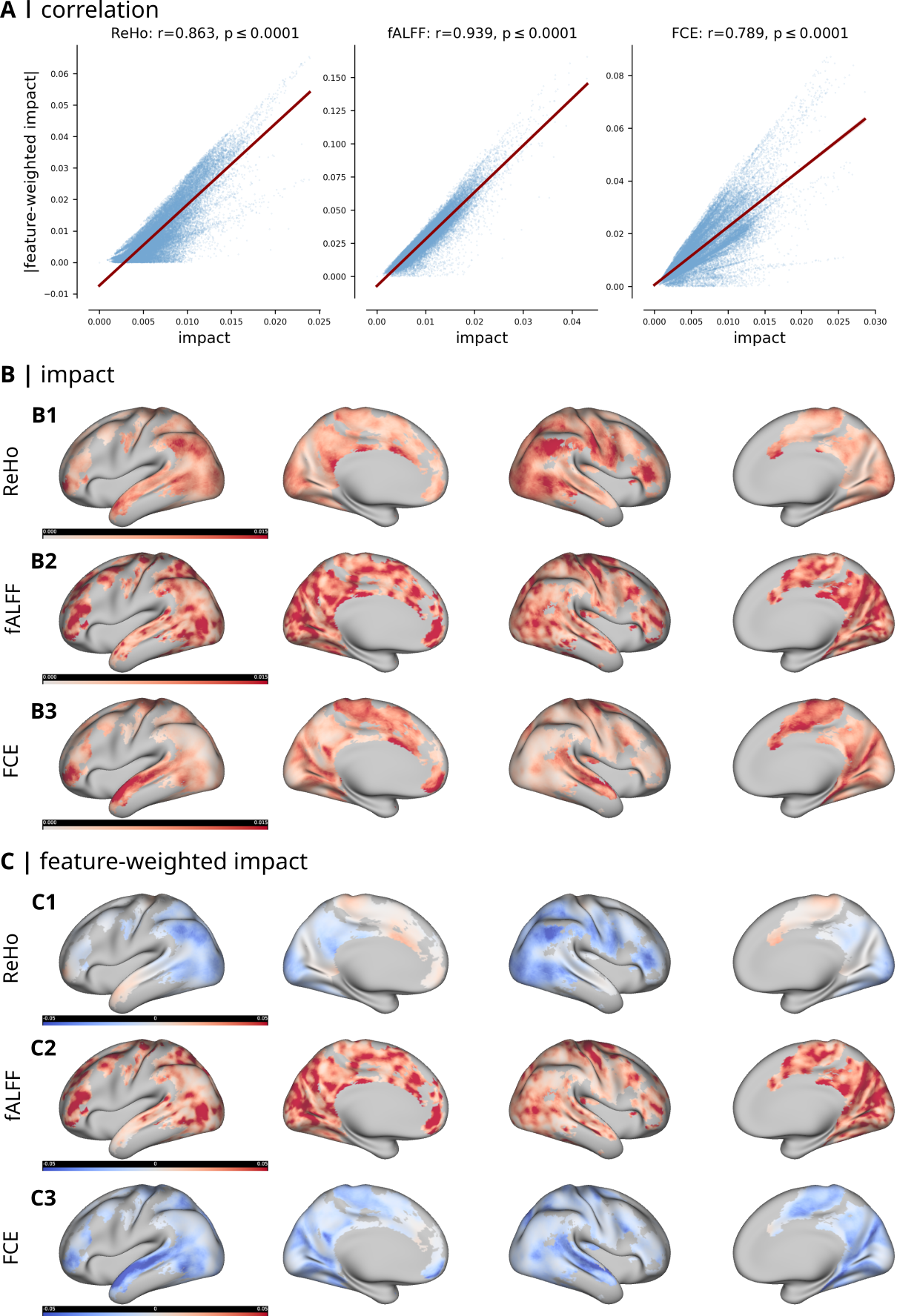} 
}
\ec 
\caption{Comparison between impact and feature-weighted impact. (A) The impact and absolute feature-weighted impact values of all three rs-fMRI metrics are highly correlated across vertices. (B) Local impact of ReHo (B1), fALFF (B2) and FCE (B3) across the cortex. (C) Local feature-weighted impact of ReHo (B1), fALFF (B2) and FCE (B3) across the cortex.}
\end{figure}


\begin{figure}[!htb]
\bc
\makebox[\textwidth][c]
{
\includegraphics[width=0.98\textwidth]{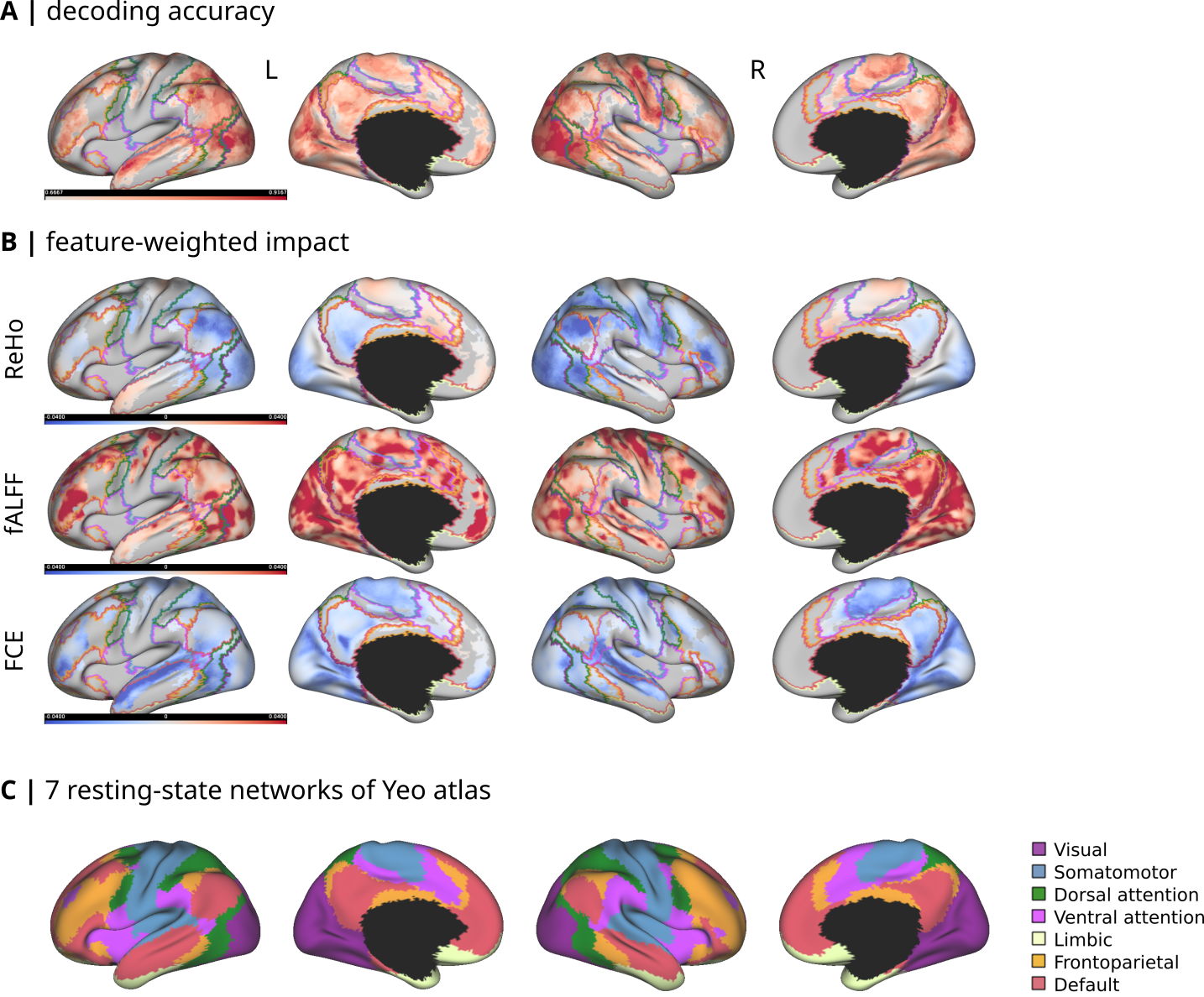} 
}
\ec 
\caption{Decoding accuracy and feature-weighted impact of FuSL overlayed with resting-state networks defined by \cite{Yeo2011_atlas}. (A) Decoding accuracy of FuSL across the cortex based on a combination of ReHo, fALFF and FCE. (B) The feature-weighted impact allows us to identify which brain areas are informative and important for the decoding and whether ReHo, fALFF and FCE are increased or decreased due to the treatment of alprazolam. (C) Displays the 7 resting-state networks defined by \cite{Yeo2011_atlas}.}
\end{figure}



\end{document}